\documentclass[longauth]{aa}
\usepackage{graphicx}
\usepackage{txfonts}
\usepackage{xcolor}
\usepackage[hidelinks]{hyperref}
\hypersetup{
  colorlinks,
  linkcolor={red},
  citecolor={blue!80!black},
  urlcolor={blue!80!black}
}
\usepackage{color}
\usepackage{siunitx}
\usepackage{mathabx}
\newcommand{\ticA}{{\rm TOI-2449}}
\usepackage[normalem]{ulem}
\usepackage{orcidlink}
\usepackage{chemformula}

\begin{document}

\title{Detection and characterisation of a 106-day transiting Jupiter : TOI-2449\,b / NGTS-36\,b
  \thanks{Based on observations collected at the European Southern Observatory under ESO programmes: 105.20GX.001, 106.21ER.001, 108.22A8.001, 108.22L8.001, 109.239V.001, 110.23YQ.001}}

\titlerunning{106-day transiting Jupiter}

\author{
  S.~Ulmer-Moll \inst{\ref{ inst1 }, \ref{geneva}, \ref{ inst3 }}\thanks{Oort Postdoctoral Fellow} \orcidlink{0000-0003-2417-7006} \and
  S.~Gill \inst{\ref{ inst4 }, \ref{ inst5 }} \orcidlink{0000-0002-4259-0155} \and
  R.~Brahm \inst{\ref{ inst6 }, \ref{ inst7 }, \ref{ inst8 }} \orcidlink{0000-0002-9158-7315} \and
  A.~Claringbold \inst{\ref{ inst5 }, \ref{ inst4 }} \orcidlink{0000-0003-1309-5558} \and
  M.~Lendl \inst{\ref{geneva}} \orcidlink{0000-0001-9699-1459} \and
  K.~Al~Moulla \inst{\ref{ia-porto}, \ref{geneva}}\thanks{SNSF Postdoctoral Fellow} \orcidlink{0000-0002-3212-5778} \and
  D.~Anderson \inst{\ref{ inst9 }} \and
  M.~Battley \inst{\ref{ inst10 }, \ref{geneva}} \orcidlink{0000-0002-1357-9774} \and
  D.~Bayliss \inst{\ref{ inst4 }} \orcidlink{0000-0001-6023-1335} \and
  A.~Bonfanti \inst{\ref{ inst11 }} \orcidlink{0000-0002-1916-5935} \and
  F.~Bouchy \inst{\ref{geneva}} \orcidlink{0000-0002-7613-393X} \and
  C.~Briceño \inst{\ref{ inst12 }} \and
  E.~M.~Bryant \inst{\ref{ inst13 }} \orcidlink{0000-0001-7904-4441} \and
  M.~R.~Burleigh \inst{\ref{ inst14 }} \orcidlink{0000-0003-0684-7803} \and
  K.~A.~Collins \inst{\ref{ inst15 }} \orcidlink{0000-0001-6588-9574} \and
  A.~Deline \inst{\ref{geneva}} \and
  X.~Dumusque \inst{\ref{geneva}} \orcidlink{0000-0002-9332-2011} \and
  J.~Eberhardt \inst{\ref{ inst16 }} \orcidlink{0000-0003-3130-2768} \and
  N.~Espinoza \inst{\ref{ inst17 }} \orcidlink{0000-0001-9513-1449} \and
  B.~Falk \inst{\ref{ inst17 }} \and
  J.~P.~Faria \inst{\ref{geneva}} \orcidlink{0000-0002-6728-244X} \and
  J.~Fernández~Fernández \inst{\ref{ inst5 }, \ref{ inst4 }} \orcidlink{0000-0002-1416-2188} \and
  P.~Figueira \inst{\ref{geneva}, \ref{ia-porto}} \orcidlink{0000-0001-8504-283X} \and
  M.~Fridlund \inst{\ref{ inst1 }, \ref{ inst19 }} \orcidlink{0000-0002-0855-8426} \and
  E.~Furlan \inst{\ref{ inst20 }} \orcidlink{0000-0001-9800-6248} \and
  M.~R.~Goad \inst{\ref{ inst14 }} \and
  R.~F.~Goeke \inst{\ref{ inst21 }} \and
  J.~Hagelberg \inst{\ref{geneva}} \and
  F.~Hawthorn \inst{\ref{ inst4 }} \orcidlink{0000-0002-8675-182X} \and
  R.~Helled \inst{\ref{ inst22 }} \orcidlink{0000-0001-5555-2652} \and
  Th.~Henning \inst{\ref{ inst16 }} \orcidlink{0000-0002-1493-300X} \and
  M.~Hobson \inst{\ref{geneva}, \ref{ inst6 }, \ref{ inst16 }} \orcidlink{0000-0002-5945-7975} \and
  S.~B.~Howell \inst{\ref{ inst23 }} \orcidlink{0000-0002-2532-2853} \and
  M.~Jafariyazani \inst{\ref{ inst24 }, \ref{ inst23 }} \orcidlink{0000-0001-8019-6661} \and
  J.~M.~Jenkins \inst{\ref{ inst23 }} \orcidlink{0000-0002-4715-9460} \and
  J.~S.~Jenkins \inst{\ref{ inst25 }, \ref{ inst26 }} \orcidlink{0000-0003-2733-8725} \and
  M.~I.~Jones \inst{\ref{eso-santiago}} \and
  A.~Jordán \inst{\ref{ inst7 }, \ref{ inst6 }, \ref{ inst8 }} \orcidlink{0000-0002-5389-3944} \and
  A.~Kendall \inst{\ref{ inst14 }} \and
  N.~Law \inst{\ref{ inst28 }} \and
  C.~Littlefield \inst{\ref{ inst29 }, \ref{ inst23 }} \orcidlink{0000-0001-7746-5795} \and
  A.~W.~Mann \inst{\ref{ inst28 }} \orcidlink{0000-0003-3654-1602} \and
  J.~McCormac \inst{\ref{ inst4 }} \orcidlink{0000-0003-1631-4170} \and
  C.~Mordasini \inst{\ref{ inst3 }, \ref{ inst30 }} \orcidlink{0000-0002-1013-2811} \and
  M.~Moyano \inst{\ref{ inst9 }} \and
  H.~Osborn \inst{\ref{ inst3 }} \orcidlink{0000-0002-4047-4724} \and
  C.~Pezzotti \inst{\ref{ inst31 }} \and
  A.~Psaridi \inst{\ref{geneva}, \ref{ inst32 }, \ref{ inst33 }} \orcidlink{0000-0002-4797-2419} \and
  S.~N.~Quinn \inst{\ref{ inst15 }} \orcidlink{0000-0002-8964-8377} \and
  T.~Rodel \inst{\ref{ inst34 }} \orcidlink{0009-0009-2175-7284} \and
  J.~E.~Rodriguez \inst{\ref{ inst35 }} \orcidlink{0000-0001-8812-0565} \and
  F.~Rojas \inst{\ref{ inst36 }, \ref{ inst6 }} \orcidlink{0000-0003-3047-6272} \and
  S.~Saha \inst{\ref{ inst25 }, \ref{ inst26 }} \orcidlink{0000-0001-8018-0264} \and
  M.~Schlecker \inst{\ref{ inst37 }} \orcidlink{0000-0001-8355-2107} \and
  S.~Seager \inst{\ref{ inst21 }, \ref{ inst38 }, \ref{ inst39 }} \orcidlink{0000-0002-6892-6948} \and
  S.~G.~Sousa \inst{\ref{ia-porto}, \ref{uporto}} \orcidlink{0000-0001-9047-2965} \and
  M.~Tala~Pinto \inst{\ref{ inst7 }, \ref{ inst6 }, \ref{ inst40b }} \orcidlink{0009-0004-8891-4057} \and
  T.~Trifonov \inst{\ref{ inst41 }, \ref{ inst42 }} \orcidlink{0000-0002-0236-775X} \and
  S.~Udry \inst{\ref{geneva}} \orcidlink{0000-0001-7576-6236} \and
  J.~I.~Vines \inst{\ref{ inst9 }} \and
  G.~Viviani \inst{\ref{ inst43 }} \and
  C.~A.~Watson \inst{\ref{ inst34 }} \orcidlink{0000-0002-9718-3266} \and
  P.~J.~Wheatley \inst{\ref{ inst5 }, \ref{ inst4 }} \orcidlink{0000-0003-1452-2240} \and
  T.~G.~Wilson \inst{\ref{ inst4 }} \orcidlink{0000-0001-8749-1962} \and
  J.~N.~Winn \inst{\ref{ inst44 }} \orcidlink{0000-0002-4265-047X} \and
  G.~Zhou \inst{\ref{ inst45 }} \orcidlink{0000-0002-4891-3517} \and
  C.~Ziegler \inst{\ref{ inst46 }} 
}

\institute{
  Leiden Observatory, Leiden University, P.O. Box 9513, 2300 RA Leiden, The Netherlands \label{ inst1 } \and
  Observatoire Astronomique de l'Université de Genève, Chemin Pegasi 51b, 1290 Versoix, Switzerland \label{geneva} \and
  Space Research and Planetary Sciences, Physics Institute, University of Bern, Gesellschaftsstrasse 6, 3012 Bern, Switzerland \label{ inst3 } \and
  Department of Physics, University of Warwick, Gibbet Hill Road, Coventry CV4 7AL, United Kingdom \label{ inst4 } \and
  Centre for Exoplanets and Habitability, University of Warwick, Gibbet Hill Road, Coventry CV4 7AL, UK \label{ inst5 } \and
  Millennium Institute for Astrophysics, Santiago, Chile \label{ inst6 } \and
  Facultad de Ingeniería y Ciencias, Universidad Adolfo Ibáñez, Av. Diagonal las Torres 2640, Peñalolén, Santiago, Chile \label{ inst7 } \and
  Data Observatory Foundation, Santiago, Chile \label{ inst8 } \and
  Instituto de Astrofísica e Ciências do Espaço, Universidade do Porto, CAUP, Rua das Estrelas, 4150-762 Porto, Portugal \label{ia-porto} \and 
  Instituto de Astronomía, Universidad Católica del Norte, Angamos 0610, 1270709, Antofagasta, Chile \label{ inst9 } \and
  Astronomy Unit, Queen Mary University of London, G.O. Jones Building, Bethnal Green, London E1 4NS, United Kingdom \label{ inst10 } \and
  Space Research Institute, Austrian Academy of Sciences, Schmiedlstrasse 6, A-8042 Graz, Austria \label{ inst11 } \and
  Cerro Tololo Inter-American Observatory, Casilla 603, La Serena, Chile \label{ inst12 } \and
  Mullard Space Science Laboratory, University College London, Holmbury St Mary, Dorking, Surrey RH5 6NT, UK \label{ inst13 } \and
  School of Physics and Astronomy, University of Leicester, Leicester LE1 7RH, UK \label{ inst14 } \and
  Center for Astrophysics $\|$ Harvard \& Smithsonian, 60 Garden Street, Cambridge, MA 02138, USA \label{ inst15 } \and
  Max-Planck-Institut für Astronomie, Königstuhl 17, 69117 Heidelberg, Germany \label{ inst16 } \and
  Space Telescope Science Institute, 3700 San Martin Drive, Baltimore, MD 21218, USA \label{ inst17 } \and
  Department of Space, Earth and Environment, Chalmers University of Technology, Onsala Space Observatory, 439 92 Onsala, Sweden \label{ inst19 } \and
  NASA Exoplanet Science Institute, Caltech/IPAC, Mail Code 100-22, 1200 E. California Blvd., Pasadena, CA 91125, USA \label{ inst20 } \and
  Department of Physics and Kavli Institute for Astrophysics and Space Research, Massachusetts Institute of Technology, Cambridge, MA 02139, USA \label{ inst21 } \and
   University of Zurich, Department of Astrophysics, Winterthurerstr. 190, CH-8057 Zurich, Switzerland \label{ inst22 } \and
  NASA Ames Research Center, Moffett Field, CA 94035, USA \label{ inst23 } \and
  SETI Institute, Mountain View, CA 94043, USA \label{ inst24 } \and
  Instituto de Estudios Astrofísicos, Facultad de Ingeniería y Ciencias, Universidad Diego Portales, Av. Ejército 441, Santiago, Chile \label{ inst25 } \and
  Centro de Astrofísica y Tecnologías Afines (CATA), Casilla 36-D, Santiago, Chile \label{ inst26 } \and
  ESO - European Southern Observatory, Av. Alonso de Cordova 3107, Vitacura, Santiago, Chile \label{eso-santiago} \and
  Department of Physics and Astronomy, The University of North Carolina at Chapel Hill, Chapel Hill, NC 27599-3255, USA \label{ inst28 } \and
  Bay Area Environmental Research Institute, Moffett Field, CA 94035, USA \label{ inst29 } \and
  Center for Space and Habitability, University of Bern, Gesellschaftsstrasse 6, 3012 Bern, Switzerland \label{ inst30 } \and
  STAR Institute, Université de Liège, Liège, Belgium \label{ inst31 } \and
  Institute of Space Sciences (ICE, CSIC), Carrer de Can Magrans S/N, Campus UAB, Cerdanyola del Valles, E-08193, Spain \label{ inst32 } \and
  Institut d'Estudis Espacials de Catalunya (IEEC), 08860 Castelldefels (Barcelona), Spain \label{ inst33 } \and
  Astrophysics Research Centre, School of Mathematics and Physics, Queen's University Belfast, Belfast BT7 1NN, UK \label{ inst34 } \and
  Center for Data Intensive and Time Domain Astronomy, Department of Physics and Astronomy, Michigan State University, East Lansing, MI 48824, USA \label{ inst35 } \and
  Instituto de Astrofísica, Facultad de Física, Pontificia Universidad Católica de Chile, Av. Vicuña Mackenna 4860, Santiago, Chile \label{ inst36 } \and
  Department of Astronomy/Steward Observatory, The University of Arizona, 933 North Cherry Avenue, Tucson, AZ 85721, USA \label{ inst37 } \and
  Department of Earth, Atmospheric and Planetary Sciences, Massachusetts Institute of Technology, Cambridge, MA 02139, USA \label{ inst38 } \and
  Department of Aeronautics and Astronautics, MIT, 77 Massachusetts Avenue, Cambridge, MA 02139, USA \label{ inst39 } \and
  Departamento de Física e Astronomia, Faculdade de Ciências, Universidade do Porto, Rua do Campo Alegre, 4169-007 Porto, Portugal\label{uporto} \and
  Department of Astronomy, The Ohio State University, 140 W. 18th Ave., Columbus, OH, 43210 \label{ inst40b } \and
  Landessternwarte, Zentrum für Astronomie der Universt\"at Heidelberg, Königstuhl 12, 69117 Heidelberg, Germany \label{ inst41 } \and
  Department of Astronomy, Sofia University St Kliment Ohridski, 5 James Bourchier Blvd, BG-1164 Sofia, Bulgaria \label{ inst42 } \and
  Institute of Physics, École Polytechnique Fédérale de Lausanne (EPFL), Observatoire de Sauverny, Chemin Pegasi 51b, 1290 Versoix, Switzerland \label{ inst43 } \and
  Department of Astrophysical Sciences, Princeton University, Princeton, NJ 08544, USA \label{ inst44 } \and
  University of Southern Queensland, Centre for Astrophysics, West Street, Toowoomba, QLD 4350 Australia \label{ inst45 } \and
  Department of Physics, Engineering and Astronomy, Stephen F. Austin State University, 1936 North St, Nacogdoches, TX 75962, USA \label{ inst46 } 
}

\date{Received 15 April 2025 / Accepted 19 September 2025}

\abstract
    { Only a handful of transiting giant exoplanets with orbital periods longer than 100 days are known.
      These warm exoplanets are valuable objects as their radius and mass can be measured
      leading to an in-depth characterisation of the planet's properties.
      Thanks to low levels of stellar irradiation and large orbital distances,
      the atmospheric properties and orbital parameters of warm exoplanets remain relatively unaltered by their host star,
      giving new insights into planetary formation and evolution.
    }
    {We aim at extending the sample of warm giant exoplanets with precise radii and masses.
      Our goal is to identify suitable candidates in the Transiting Exoplanet Survey Satellite (TESS) data and
      perform follow-up observations with ground-based instruments.}
    {We use the Next Generation Transit Survey (NGTS) to detect additional transits of planetary candidates
      in order to pinpoint their orbital period.
      We also monitored the target with several high-resolution
      spectrographs to measure the planetary mass and eccentricity.
      We studied the planet's interior composition with 
      a planetary evolution code to determine the planet's metallicity.
    }
    { We report the discovery of a 106-day period Jupiter-sized planet around the G-type star \ticA\ / NGTS-36.
      We jointly modelled the photometric and radial velocity data
      and find that the planet has a mass of $\rm 0.70^{+0.05}_{-0.04}\,M_{J}$ and a radius of $\rm 1.001\pm0.009\,R_{J}$.
      The planetary orbit has a semi-major axis of 0.449 au and is slightly eccentric ($\rm e=0.098^{+0.028}_{-0.030}$). %($\rm e = 0.098$). 
      We detect an additional 3-year signal in the radial velocity data likely due to the stellar magnetic cycle.
      Based on the planetary evolution models considered here, we find that 
      \ticA\,b / NGTS-36\,b contains $11^{+6}_{-5}$\,M$_{\oplus}$ of heavy elements
      and has a marginal planet-to-star metal enrichment of $3.3^{+2.5}_{-1.8}$. 
      Assuming a Jupiter-like bond albedo, \ticA\,b / NGTS-36\,b has an equilibrium temperature of 400\,K
      and is a good target for understanding nitrogen chemistry in cooler atmospheres.
    }
    {}


  \maketitle

  \section{Introduction}

  Gas giant exoplanets play a crucial role in the formation and evolution of planetary systems. 
  They form within the first 10 Myrs, before the gas disc dissipates (e.g. \citealt{mordasini_2024}), and contain the majority of the mass of a planetary system.
  Gas giants are known to shape the architecture of planetary systems (e.g. \citealt{levison_2003,kong_2024}),
  but the exact role of giant planets and the mechanisms driving their formation and migration remain to be understood.

  The discovery of close-in giant exoplanets, with orbital periods of a few days (e.g. \citealt{mayor_1995a}),
  brought more questions with regards to their formation and migration pathways. Two types of models are usually considered :
  the in-situ formation (e.g. \citealt{bodenheimer_2000}) and the ex-situ formation, where these gas giants must have formed at
  large distances from the host stars and migrated inwards to be found at their current location (e.g. \citealt{lin_1996}).
  Several migration scenarios are needed to explain the range of orbital periods and eccentricities at which these giant planets are found.
  Generally, the two main channels envisioned are high-eccentricity migration (e.g. \citealt{rasio_1996}) and disc migration (e.g. \citealt{goldreich_1980,baruteau_2014} for a review).

  Studying the orbital architecture and in particular stellar obliquities (spin-orbit angles)
  of close-in giant planets are key to understanding which migration scenarios are needed to reproduce the observed population .
  Transiting exoplanets are excellent probes to determine projected spin-orbit angles,
  often measured with observations of the Rossiter-McLaughin effect.

  Unfortunately the orbital and physical parameters of hot Jupiters, planets with orbital period shorter than 10 days,
  are largely influenced by the interactions with their host star, which can change their original properties.
  Hot Jupiters are subject to strong tidal interactions with their host star, which dampens their orbital eccentricity
  and could dampen stellar obliquities (e.g. \citealt{albrecht_2012,albrecht_2022a}).
  Hot Jupiters also have larger radii than expected from
  interior structure modelling due to their proximity with the host star and the intense stellar irradiation they receive.
  \cite{dawson_2018} review the mechanisms leading to the inflation of hot Jupiter radii.

  Contrary to hot Jupiters, warm Jupiters, usually defined as planets with orbital periods between 10 and 200 days (e.g. \citealt{dawson_2018,wang_2024}),
  are less susceptible to stellar interactions and likely retain their post formation and migration characteristics.
  The sample of well characterised warm Jupiters is still relatively small :
  90 warm giant planets have precise masses and radii (uncertainty less than 25\% and 8\% respectively)\footnote{\href{https://exoplanetarchive.ipac.caltech.edu/cgi-bin/TblView/nph-tblView?app=ExoTbls&config=PS}{NASA exoplanet archive} queried on April, 15, 2025 for planets with masses larger than 0.1 $\rm M_{J}$}. 
  Since its launch in December 2018, the TESS satellite \citep{ricker_2015a} has been conducting a near all sky survey with 27-day long pointings.
  Towards the poles, the pointings overlap to provide an uninterrupted observation of up to one year.
  Most transiting warm Jupiters appear as a single transit event in a given pointing.
  \citet{villanueva_2019a} and \citet{cooke_2019} pointed out the need to identify single or duo transit events in order to discover long-period planets and published predicted yields.
  From simulations including TESS Year 1 and 3 light curves, \citet{rodel_2024} found that about 110 planets
  with orbital periods larger than 100 days should be detected and that 75\% of those will be monotransits.

  Long-term radial velocity campaigns enabled mass and eccentricity measurements for these transiting planets,
  leading to well-characterised systems (e.g. \citealt{schlecker_2020,hobson_2021,eberhardt_2023,ulmer-moll_2023,brahm_2023}).
  Photometric follow-up campaigns are also required to recover the orbital period of warm transiting planets.
  Citizen scientists are involved in this effort both in the identification of single transit events
  (e.g. TOI-2180\,b, a 260-day planet; \citealt{dalba_2022}) and the photometric observations of candidates (e.g. \citealt{sgro_2024}).
  The Next Generation Transit Survey (NGTS; \citealt{wheatley_2018}) has a multi-year monitoring program of single and duo transit candidates which has led
  to the discovery of several transiting warm Jupiters with orbital periods ranging from just above
  20 days to up to 100 days (e.g. \citealt{gill_2020a, ulmer-moll_2022a, battley_2024}).

  In this context, we report the discovery of a transiting warm Jupiter orbiting \ticA\, characterised with TESS and NGTS
  data and radial velocity data.
  In Section~\ref{observations} we present the photometric data, radial velocities, and high-resolution imaging data.
  The methods for the determination of the stellar and planetary parameters are detailed in Section~\ref{methods}.
  In Section~\ref{results}, we present the results of the joint photometric and radial velocity fit and we determine the
  heavy element mass of \ticA\,b. Finally, we discuss and summarize our work in Sections~\ref{sec:discussion} and \ref{conclusion}.

\section{Observations}
\label{observations}

\ticA\, was observed by TESS (Section~\ref{sec:tess}) and
follow-up photometric observations were obtained with NGTS (Section~\ref{sec:ngts}).
Ground-based spectroscopic observations were carried out with the high-resolution spectrographs
CHIRON, CORALIE, FEROS, and HARPS which are presented in Sections~\ref{sec:chiron}, \ref{sec:coralie}, \ref{sec:feros}, and \ref{sec:harps} respectively.
High-resolution imaging data and results are detailed in Sections~\ref{sec:zorro} and ~\ref{sec:soar}.

\subsection{TESS photometry}
\label{sec:tess}

The Transiting Exoplanet Survey Satellite (TESS; \citealt{ricker_2015a}) is a photometric survey mission which scans almost the entire sky.
\ticA\ was observed by TESS in sectors 4 and 5 (from 2018 Oct 19 to 2018 Dec 11) at two-minute cadence and in sectors 31 and 32
(from 2020 Oct 22 to 2020 Dec 16) at a cadence of 10 minutes.

\ticA\,b was first announced as a TESS Object of Interest (TOI) on January 6, 2021 with a single transit event detected in sector 31 by the Quick Look Pipeline (QLP) at MIT \citep{huang_2020,huang_2020a}.
We obtained the light curves from the four TESS sectors through the Mikulski Archive for Space Telescopes
(MAST) archive\footnote{\href{https://mast.stsci.edu/portal/Mashup/Clients/Mast/Portal.html}{mast.stsci.edu}}.
We downloaded the outputs of the Science Processing Operation Center (SPOC; \citealt{jenkins_2016}) pipeline
and we used the pre-search data conditioning simple aperture photometry (PDCSAP; \citealt{stumpe_2012,smith_2012,stumpe_2014}) in our analysis.
The image data were reduced and analysed by the SPOC at NASA Ames Research Center.
The PDCSAP light curves were corrected for dilution and for systematic trends shared by other stars on the detector.

We checked for contaminant sources in the aperture with the \texttt{tpfplotter} software package \citep{aller_2020}. 
No neighbouring sources were found down to a G magnitude difference of 8 for the four TESS sectors.
The two nearest neighboring stars have flux ratios of 0.0032 and 0.0009 in the G band.
We include the figure of the TESS target pixel file for sector 31 in Appendix~\ref{appendix:tpf_plot}.

\subsection{NGTS photometry}
\label{sec:ngts}

The Next Generation Transit Survey operates a set of twelve robotic telescopes located at Cerro Paranal, Chile.
Each unit is a 20 cm telescope ﬁtted with back-illuminated deep-depletion CCD cameras \citep{wheatley_2018}.
NGTS uses a custom filter ranging from 520 to 890 nm.
For TOI-2449, with a T-band magnitude of 9.9, the expected photometric precision for a single NGTS telescope is 400 ppm per 30 mins.
At T-band magnitude brighter than 9.5, the 12 NGTS telescopes reach a photometric precision of 100 ppm per 30 mins (e.g. \citealt{bryant_2020,obrien_2021}).

After the release of the new transiting candidate around \ticA\, the target was included in
the on-going NGTS program aimed at the discovery of warm and temperate exoplanets.
\ticA\ was monitored for a first campaign
which resulted in 80 nights of observations with one NGTS camera from 2021-Jan-15 to 2021-Apr-23.
No transit event was detected in this first NGTS campaign.
The second campaign consisted of 94 nights of observations with one NGTS camera from 2021-Jul-05 to 2021-Dec-03.
One partial transit matching the transit depth of the transiting event in TESS was identified on 2021-Sept-14.
After this detection, the maximum possible period was 318.4 days and five period aliases (all above 50 days) were still compatible with the TESS and NGTS data.
A second partial transit was observed on 2022-Nov-13  with six cameras based on the radial velocity solution.
This third transit reduced the set of possible orbital periods to only two period aliases: 53.07 and 106.14 days.
During the first transit observation, the photometric precision was 380 ppm per 30 mins. During the second transit observations the 30 min photometric precision was between 530 and 680 ppm per camera. The seeing and wind speed were higher during this second observation and probably impacted the photometric precision.
The NGTS data shows a long-term decreasing trend which cannot be attributed to instrumental effects and is likely due to long-term stellar activity.
The nightly binned NGTS data are presented in Fig.~\ref{fig:all_ngts} along with three selected nights showing light curves
with and without a transit detection.

\subsection{CORALIE spectroscopy}
\label{sec:coralie}

We used the CORALIE spectrograph \citep{baranne_1996a,queloz_2001a,segransan_2010} to first confirm the planetary nature
of \ticA\,b and then monitor the radial velocity of the target for two years.
CORALIE is a high-resolution fiber-fed spectrograph installed at the Swiss Euler 1.2m
telescope\footnote{\href{https://www.eso.org/public/teles-instr/lasilla/swiss/}{eso.org/public/teles-instr/lasilla/swiss}} located in La Silla, Chile.
CORALIE has a spectral resolution of 60,000 with a sampling per resolution element equal to 3 pixels. CORALIE is fed by two fibers.
The first fiber is used for the science observation. The second fiber is used either to observe a Fabry-Pérot étalon for drift calibration purposes
or to observe the sky for background subtraction. We observed \ticA\ together with the simultaneous Fabry–Pérot étalon.

We obtained 32 CORALIE spectra of \ticA\ from 2021-Jan-17 to 2023-Mar-11 with exposure times varying between 900 and 1800 seconds.
The objectives of the radial velocity observations are to rule out spectroscopic binaries with the first two CORALIE measurements
and then to detect the reflex motion induced by the transiting candidate with a monitoring over several months with a radial velocity precision of about $\rm 10\,m\,s^{-1}$.
The spectra were reduced with the standard data reduction pipeline and cross-correlated with a G2-type stellar mask to obtain the radial velocity measurements
and associated errors (e.g. \citealt{pepe_2002a}). The radial velocities are presented in Table \ref{table:table_rvs}. Along with the radial velocities,
the pipeline provides useful parameters which can be used as stellar activity indicators: the full width half maximum (FWHM), contrast,
and bisector inverse slope (BIS; \citealt{queloz_2001b}) of the cross-correlation function (CCF).

\subsection{CHIRON spectroscopy}
\label{sec:chiron}

CHIRON is a fiber-bed high-resolution spectrograph installed at the 1.5m telescope located
in Cerro Tololo INter-american Observatory, Chile (CTIO; \citealt{tokovinin_2013}).
CHIRON monitored \ticA\ for one year, from 2021-Jan-08 to 2022-Jan-29, collecting 72 observations
with an exposure time varying between 900 and 1200 seconds, leading to a signal-to-noise ratio
per extracted pixel between 17 and 40, depending on the atmospheric conditions and airmass.
Spectral extraction and calibration was performed via the standard CHIRON pipeline \citep{paredes_2021}.
Line profiles and radial velocities are determined from each spectrum via a least-squares deconvolution between
the observed spectrum and an synthetic non-rotating template, generated from ATLAS9 model atmospheres \citep{castelli_2003}.
Radial and line broadening velocities are determined by modeling the line profiles with an analytic broadening kernel
that encompasses the effects of rotational, macroturbulence, and instrumental broadening as per \citet{gray_2005}.
The radial velocity measurements are reported in Table \ref{table:table_rvs}.

\subsection{FEROS spectroscopy}
\label{sec:feros}

\ticA\ was also monitored by the Fiber-fed Extended Range Optical Spectrograph (FEROS), located at the 2.2 m telescope in La Silla, Chile.
FEROS has a resolution of 48,000 and covers the optical range from 360 to 920 nm \citep{kaufer_1999}.
\ticA\ was observed over two years cumulating 17 observations under the programs: 0106.A-9014, 0107.A-9003, 0108.A-9003, 0109.A-9003.
The observations were done using the simultaneous wavelength calibration mode with an average exposure time of 300 seconds,
leading to spectra with typical signal-to-noise ratio per resolution element between 60 and 90.
The data were reduced with the \texttt{CERES} pipeline \citep{brahm_2017} which provides the radial velocity measurement and associated error as
well as the FWHM and BIS of the CCF. The radial velocity measurements are listed in Table \ref{table:table_rvs}.

\subsection{HARPS spectroscopy}
\label{sec:harps}

HARPS is a high resolution fiber-fed spectrograph installed at the European Southern Observatory (ESO)
3.6 m telescope in La Silla, Chile \citep{mayor_2003a}. Given that CORALIE, CHIRON, and FEROS observations hinted at the detection of a planet around \ticA,
additional observations were acquired with HARPS in order to refine the orbital solution and planetary
parameters with a radial velocity precision at the $\rm m\,s^{-1}$ level.
During two years, we used the HARPS spectrograph to monitor
\ticA\ with the high-accuracy mode and exposure times varying between 900 and 1200 seconds depending on the weather conditions.
\ticA\ was observed for a total of 55 observations under several program aiming to confirm transiting warm Jupiters:
105.20GX.001, 106.21ER.001, 108.22L8.001, 108.22A8.001, 109.239V.001, 110.23YQ.001.
The HARPS data were reduced with the standard data reduction pipeline and the radial velocities were obtained
with the cross-correlation technique using a G2 mask. The radial velocity measurements are reported in Table \ref{table:table_rvs}.
The HARPS spectra were co-added and then used to perform the stellar analysis in Section~\ref{stellar-analysis}.

\subsection{Zorro imaging}
\label{sec:zorro}

As part of the validation and confirmation process for a transiting exoplanet observation,
high-resolution imaging is one of the critical assets required. The presence of a close companion star,
whether truly bound or along the line of sight, will provide additional light contamination of the observed transit,
leading to derived properties for the exoplanet and host star that are incorrect \citep{ciardi_2015,furlan_2017,furlan_2020}.
Given that nearly one-half of FGK stars are in binary or multiple star systems \citep{matson_2018}
high-resolution imaging yields crucial information toward our understanding of each discovered exoplanet
as well as more global information on exoplanetary formation, dynamics and evolution \citep{howell_2021}.

TOI-2449 was observed on 2021-Sept-22 using the Zorro speckle instrument on Gemini South \citep{scott_2021}.
Zorro provides simultaneous speckle imaging in two bands (562 nm and 832 nm) with output data products including
a reconstructed image and robust contrast limits on companion detections \citep{howell_2011}.
Seven sets of 1000 $\times$ 0.06 sec exposures were collected and subjected to Fourier analysis
using the Zorro standard data reduction pipeline.
Figure~\ref{fig:zorro} shows the resulting $5\,\sigma$ contrast curves
and the 832 nm reconstructed speckle image. We find that TOI-2449 is a single star
with no close companion brighter than 5 to 7 magnitudes from the diffraction limit (~20 mas) out to 1.2”.
At the distance of TOI-2449, these angular limits correspond to spatial limits of 3 to 188 au.

\subsection{SOAR imaging}
\label{sec:soar}
We also searched for stellar companions to TOI-2449 with speckle imaging on the 4.1-m Southern Astrophysical Research (SOAR) telescope \citep{tokovinin_2018} on 2021-Feb-27, observing in Cousins I-band, an overlapping visible bandpass as TESS. This observation was sensitive to a 5.8-magnitude fainter star at an angular distance of 1 arcsec from the target. More details of the observation are available in \citet{ziegler_2020}. The $5\,\sigma$ detection sensitivity and speckle auto-correlation functions from the observations are shown in Fig.~\ref{fig:soar}. No nearby stars were detected within $3\arcsec$ of TOI-2449 in the SOAR observations.

\begin{table}
  \caption{Radial velocities of \ticA.}
  \label{table:table_rvs}
  \begin{tabular}{l l l l}
    \hline
    \hline
    \noalign{\smallskip}
    Time             & RV                   & RV error                  & Instrument\\
    BJD              & [$\rm kms^{-1}$]      & [$\rm kms^{-1}$]          & \\
    \hline
    \noalign{\smallskip}
    2459222.61066 & -4.823 & 0.022 & CHIRON \\
    2459223.61061 & -4.864 & 0.022 & CHIRON \\
    2459225.61142 & -4.815 & 0.023 & CHIRON \\
    ...&&&\\
    2460294.61190 & -3.404 & 0.010 & CORALIE \\
    2460330.64524 & -3.433 & 0.017 & CORALIE \\
    2460359.54398 & -3.448 & 0.008 & CORALIE \\
    \hline
  \end{tabular}
  \tablefoot{Full table is available at CDS.}
\end{table}

\section{Methods}
\label{methods}

\subsection{Stellar parameter determination}
\label{stellar-analysis}

We used the ARES+MOOG method described by \citealt[][]{Sousa-21, Sousa-14, Santos-13} to obtain the stellar spectroscopic parameters ($T_{\mathrm{eff}}$, $\log g$, microturbulence, [Fe/H]). The equivalent widths (EW) were measured using the ARES code\footnote{The last version, ARES v2, can be downloaded at \url{https://github.com/sousasag/ARES}} \citep{Sousa-07, Sousa-15}. For this spectral analysis we used a combined HARPS spectrum of TOI-2449 (SNR = 250) to measure the equivalent widths for the list of lines presented in \citet[][]{Sousa-08}. The best set of spectroscopic parameters for each spectrum was found by using a minimization process to find the ionization and excitation equilibrium. This process makes use of a grid of Kurucz model atmospheres \citep{Kurucz-93} and the latest version of the radiative transfer code MOOG \citep{Sneden-73}. We also derived a more accurate trigonometric surface gravity (adopted value) using recent Gaia data.
We matched the star with the Gaia ID in DR3 using their coordinates and the VizieR catalogues (I/355/gaiadr3, \citealt{gaiacollaboration_2023}).
We followed the same procedure as described in \citet[][]{Sousa-21} which provided a consistent value when compared with the spectroscopic surface gravity.

To determine the stellar radius of TOI-2449, we utilised a MCMC modified infrared flux method \citep[IRFM --][]{Blackwell1977,Schanche2020}. Within this approach we computed synthetic photometry from constructed spectral energy distributions (SEDs) built from two atmospheric model catalogues \citep{Kurucz1993,Castelli2003} that were constrained using our stellar spectral results. These synthetic fluxes were compared to broadband observations in the following passbands: Gaia $G$, $G_\mathrm{BP}$, and $G_\mathrm{RP}$, 2MASS $J$, $H$, and $K$, and WISE $W1$ and $W2$ \citep{Skrutskie2006,Wright2010,GaiaCollaboration2022} to derive the stellar bolometric flux. From this we obtained the stellar effective temperature and angular diameter that was converted into the stellar radius using the offset-corrected Gaia parallax \citep{Lindegren2021}. We conducted a Bayesian model averaging of the radius posterior distribution from the individual atmospheric catalogues to account for model uncertainties.

The stellar effective temperature, metallicity and radius constitute the basic input set we used to derive the stellar mass $M_{\star}$ and age $t_{\star}$ with two different stellar evolutionary codes. We first used the isochrone placement algorithm \citep{bonfanti2015,bonfanti2016} and its capability of interpolating the input parameters within pre-computed grids of PARSEC\footnote{\textsl{PA}dova and T\textsl{R}ieste \textsl{S}tellar \textsl{E}volutionary \textsl{C}ode: \url{http://stev.oapd.inaf.it/cgi-bin/cmd}} v1.2S \citep{marigo2017} isochrone and tracks. As detailed in \citet{bonfanti2016}, to enhance convergence we further inputted $v\sin{i_{\star}}$ as a proxy for the stellar rotation period and benefited from the synergy between the isochrone fitting and the gyrochronological relation from \citet{barnes2010} to get a first pair of mass and age estimates. A second pair of estimates, instead, was derived by the CLES \citep[Code Liègeois d'Évolution Stellaire;][]{scuflaire2008} code that generates the best-fit stellar evolutionary track following a Levenberg-Marquadt minimisation scheme constrained by the basic set of input parameters \citep[see][]{salmon2021}.
The two stellar evolutionary codes led to stellar masses of $M_{\star}=1.065\pm0.049\,M_{\odot}$ with PARSEC and $M_{\star}=1.091\pm0.041\,M_{\odot}$ with CLES, and stellar ages of $t_{\star}=2.8\pm1.3$ Gyr with PARSEC and $t_{\star}=2.1\pm1.5$ Gyr with CLES.
After checking the mutual consistency of the two respective pairs of outcomes with the $\chi^2$-based criterion outlined in \citet{bonfanti2021},
we combined the two mass and age distributions and we obtained $M_{\star}=1.079_{-0.048}^{+0.045}\,M_{\odot}$ and $t_{\star}=2.5_{-1.5}^{+1.4}$ Gyr \citep[see][for further details]{bonfanti2021}.
The results are presented in Table~\ref{table:stellar-params}.

\begin{table}
  \caption{Stellar properties and stellar parameters.}
  \label{table:stellar-params}
  \centering
  \begin{tabular}{l c c}
    \hline
    \hline
    \noalign{\smallskip}
    & \ticA\  & \\
    \hline
    \noalign{\smallskip}
    Other Names & & \\
    \noalign{\smallskip}
    2MASS     & J04315623-3427189                        & 2MASS\\
    Gaia      & 4870809920906672384                      & Gaia\\
    TYC       & 7041-01581-1                             & Tycho\\
    TESS      & TIC\,170729775                           & TESS\\
    TOI       & \ticA                                    & TESS\\
    
    \hline
    \noalign{\smallskip}
    Astrometric Properties & & \\
    \noalign{\smallskip}
    R.A. (epoch 2015.5)                  & $\rm 04^{h}31^{m}56^{s}.25$           & TESS\\
    Dec  (epoch 2015.5)                  & $\rm \text{-}34^{\circ}27'19{''}.92$  & TESS\\
    $\mu$R.A.($\rm mas yr^{-1}$)          & 10.076 ± 0.013                      & Gaia DR3\\
    $\mu$Dec.($\rm mas yr^{-1}$)          & -58.496 ± 0.016                     & Gaia DR3\\
    Parallax (mas)                       & 6.451±0.013                         & Gaia DR3\\
    Distance (pc)                        & 155.01±0.31                         & Gaia DR3\\
    \hline
    \noalign{\smallskip}
    Photometric Properties & & \\
    \noalign{\smallskip}
    V (mag)          & 10.416±0.005                           & Tycho\\
    B (mag)          & 11.003±0.077                           & Tycho\\
    G (mag)          & 10.2924±0.0003                         & Gaia\\
    T (mag)          & 9.899±0.006                           & TESS\\
    J (mag)          & 9.407±0.026                            & 2MASS\\
    H (mag)          & 9.149±0.027                            & 2MASS\\
    Ks(mag)          & 9.073±0.025                            & 2MASS\\
    \hline
    \noalign{\smallskip}
    Bulk Properties                        &                  & \\
    \noalign{\smallskip}
    $T_{eff}$ (K)                       & $6021 \pm 62$    & Sec.~\ref{stellar-analysis} \\
    log g ($\rm cms^{-2}$)                  & $4.40 \pm 0.02$   & Sec.~\ref{stellar-analysis} \\
    $\rm V_t$ micro ($\rm kms^{-1}$)        & $ 1.13 \pm 0.02$ & Sec.~\ref{stellar-analysis} \\
    $\rm [Fe/H]$ (dex)                     & -$0.03 \pm 0.04$    & Sec.~\ref{stellar-analysis} \\
    $\rm v.sini$ ($\rm km.s^{-1}$)          & $5.3 \pm 0.5$    & Sec.~\ref{stellar-analysis}\\
    Age (Gyr)                              & $2.5_{-1.5}^{+1.4}$      & Sec.~\ref{stellar-analysis} \\
    Radius ($R_\odot$)                      & $1.065 \pm 0.007$      & Sec.~\ref{stellar-analysis} \\
    Mass ($M_\odot$)                        & $1.079_{-0.048}^{+0.045}$ & Sec.~\ref{stellar-analysis} \\

    \hline
  \end{tabular}
  \tablebib{2MASS \cite{skrutskie_2006}; GAIA EDR3 \cite{gaiacollaboration_2021}; Tycho \citep{hog_2000}.}
\end{table}

\subsection{Radial velocity analysis}

\subsubsection{Search for periodic signals}
\begin{figure}
  \includegraphics[width=0.95\hsize]{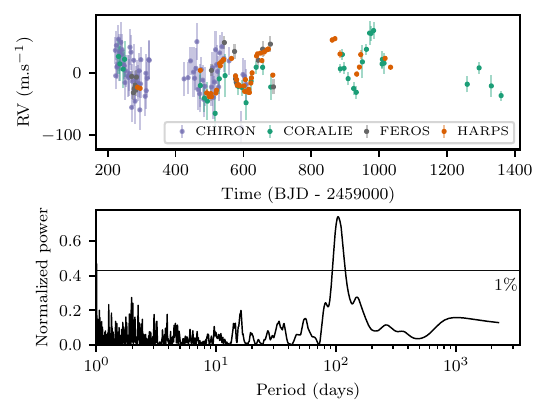}
  \caption{Radial velocities of \ticA. Top: Radial velocity data spans 1136 days (CHIRON : blue, CORALIE: green, FEROS : grey, HARPS: orange).
    Bottom: Generalised Lomb-Scargle periodogram of the radial velocities. The black line marks the 1\% false alarm probability.}
  \label{fig:rvs_plot}
\end{figure}

As described in Section~\ref{observations}, following the announcement of transiting candidate \ticA\,b,
the host star was observed by four high-resolution spectrographs covering a time span of 1136 days.
The radial velocity data were first analysed to search for the signal of the transiting planet and potential additional planets.
The full dataset of radial velocities and the generalised Lomb-Scargle periodogram are presented in Fig.~\ref{fig:rvs_plot}.
This first periodogram shows a clear detection of a periodic signal with a wide peak around 106 days. This periodicity is in line with the period alias at 106.14 days of the transiting candidate found in the TESS and NGTS data. There is no significant periodic signal around the other period alias at 53 days, thus we associate the photometric detection and this radial velocity signal to originate from the same object : a transiting planet at 106 days.
To search for additional signals, we subtracted the best fit sinusoid with the period corresponding to the highest power (106 days) from the radial velocity data. We then re-computed the periodogram and selected the next highest peak which exceeds the 1\% false alarm probability (FAP). We repeated this procedure until no peak is detected above the 1\% FAP. We find a second significant signal at around 1100 days, which is similar to the time span of the data, and a third significant signal at 34 days.
Figure~\ref{fig:rv_periodo} displays the periodogram in period space of the whole radial velocity dataset, the activity indicators, and the periods mentioned previously are highlighted. The activity indicators are computed from the CORALIE, FEROS, and HARPS spectra only. We note that some spectra are too low signal-to-noise to compute reliable values.

We ran a radial velocity only analysis to determine if a model with three planetary signals is the preferred one. We used \texttt{kima} \citep{faria_2018a}, a software package which allows one to fit for the number of planets as a free parameter, along with all the parameters describing the Keplerian orbits, instruments offsets and jitters. We used a uniform prior on the number of planets from 0 to 3 planets. The priors for the other parameters are detailed in Table~\ref{table:prior-kima} and in particular we note that we set a LogUniform prior for the orbital period ranging from 1 to 3000 days (slightly more than 2.5 times the span of the radial velocity data). After running our model with \texttt{kima}, we have an effective sample size (ESS) of 22790, with no samples for the no-planet nor the one-planet model. We find that a two-planet model is largely favoured over all the other models by comparing the probabilities : the Bayes factor of the two-planet versus the one-planet model is thus larger than the ESS of 22790, largely above the usual threshold of 150 cited for a decisive detection (e.g. \citealt{feroz_2011}). The Bayes factor of the three-planet model versus the two-planet model is only of 0.14 demonstrating that there is not enough evidence to claim the detection of a third signal.

\begin{figure}
  \includegraphics[width=0.95\hsize]{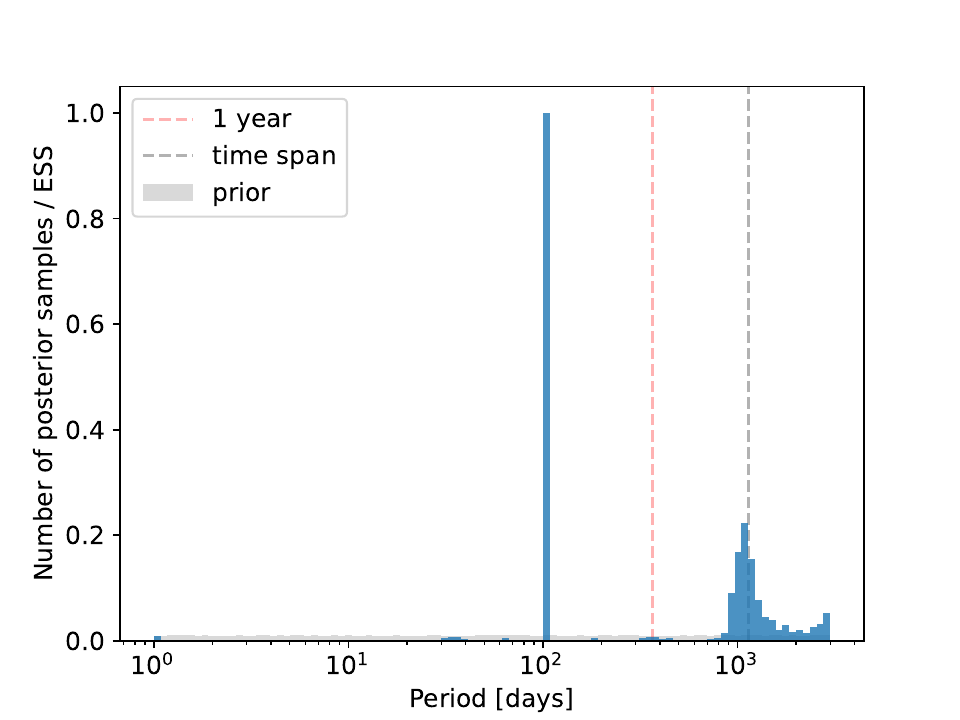}
  \caption{Posterior distribution of the orbital period for a model with up to three planets (blue histogram).
    The prior on the orbital period is identical for the three planets and shown in grey. The effective sample size (ESS) is equal to 22790 samples.}
  \label{fig:kima_periods}
\end{figure}

Only two signals are confidently detected when modelled as Keplerian functions with \texttt{kima}. The posterior distribution of the orbital periods is shown in Fig.~\ref{fig:kima_periods} displaying two peaks close to 106 and 1140 days.
In the periodogram search, the 34-day signal has a relatively high probability (about 1\%) to be a false alarm when compared to the first two signals detected : FAPs are $10^{-52}$ and $10^{-16}$ for the 106 and 1100-day signals respectively. Hence the 34-day signal is less significant and could be due to the sensitivity of the periodogram to the choice of noise model (e.g. \citealt{delisle_2020}). It is also worth mentioning that in the \texttt{kima} analysis the orbital period priors are wide (see Table~\ref{table:prior-kima}) which may result in a conservative Bayes factor values when comparing the two and three-planet models.

We note that the 34-day signal originally found in the periodogram search could be explained as an alias of the stellar rotation period. Using the log $\rm R'_{HK}$, we estimated the stellar rotation to be $16\pm3$ days (Eq. 3 and 4 from \citealt{noyes_1984}). We computed the periodogram of the NGTS data and we find hints of a 16-day periodic signal in the second season of data but not in the first one.
For the first signal, we retrieved an orbital period of $106.06^{+0.52}_{-0.55}$ days which is in agreement with one period alias of the transiting object (106.143 days) within $\rm1\,\sigma$. We independently detected the reflex motion of the transiting candidate thus confirming the detection of a 106-day planet around \ticA. For the second signal, the posterior distribution of the orbital period is less constrained and results in an orbital period of $1140^{+588}_{-138}$ days.
While the origin of the first signal is clear and in line with the transiting candidate detected in the photometric data,
the origin of the second signal with a period of about 3 years remains to be determined.

\subsubsection{Possible origin of the 3-year signal}
\label{sec:3year-signal}

We investigated if the signal at $1140^{+588}_{-138}$ days could be associated with the stellar magnetic cycle.
\citet{suarezmascareno_2016} report in their combined study of 150 stars the statistics on the length of known magnetic cycles. Based on 55 G-type stars, the authors find that the distribution of cycle length peaks between 2 and 4 years and then slowly decrease until 12 years. They also report that the median length of a cycle is $6\pm3.6$ years. Regarding the stellar rotation period, the G stars in their sample have a median rotation period of $18.4\pm11.1$ days. These values are well in agreement with what we measured for \ticA : a G0/G1-type star with a stellar rotation period of $16\pm3$ days and a potential magnetic cycle of $3.1^{+1.6}_{-0.4}$ years. 

Moreover, \citet{isaacson_2024} report the search for activity cycles in 285 stars from the California Legacy survey. The authors detect activity cycles for 138 stars in their sample. For the 29 G-stars with detected activity cycles, they find that the periods of the activity cycle range from 3.9 to 23 years and the stellar rotation periods range from 14 to 34 days. Interestingly, they find tightly constrained cycle periods for stars with $\rm log(R'_{HK})$ between -4.7 and -4.9. For the more active G-stars, the cycle periods are equal to or smaller than 5 years, while less active G-stars have cycle period ranging from 5 to 23 years. With a $\rm log(R'_{HK})$ of -4.9, an activity cycle of about 3 years for \ticA\ is coherent with this recent study.

\begin{figure}
  \includegraphics[width=\hsize]{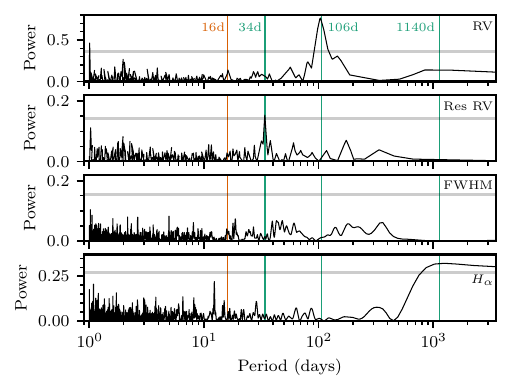}
  \caption{Generalised Lomb-Scargle periodogram of the radial velocities, the radial velocity residuals after joint modelling (see Section~\ref{sec:joint_fit}), FWHM of the CCF, and $\rm H_{\alpha}$ indicator. Vertical lines highlight the relevant periods (red : stellar rotation, green : significant signals). The grey line marks the 1\% false alarm probability.}
  \label{fig:rv_periodo}
\end{figure}

The hypothesis of a 3-year magnetic cycle is supported by the study of the stellar activity indicators derived from the high-resolution spectra of \ticA. 
In Fig.~\ref{fig:rv_periodo} looking first at the FWHM of the CCF, 
we find no significant periodicity at 106 days and there is no significant signal close to 1100 days.
One indicator has a significant peak close to 1140 days : $H_\alpha$ with a periodicity around 1200 days.
With their work on the Sun as a star, \cite{livingston_2007} report that CaII K tracks very well the solar magnetic cycle
and that other chrosmospheric lines such as $H_\alpha$ trace the same variations as Ca II K. For FGK stars,
Ca II and $H_\alpha$ are also shown to trace long-term activity cycles as well as short-term signals (e.g. \citealt{gomesdasilva_2022}).
Hence, we propose that this 3-year signal is likely emerging from stellar activity and can be attributed to the magnetic cycle of \ticA.

\subsubsection{Stellar activity modelling}

We aim at choosing the best model for the 3-year signal in the radial velocity data. Using \texttt{kima}, we first modelled our dataset with only the transiting planet \ticA\,b, setting the prior on the orbital period to a normal distribution around the value obtained from a preliminary fit of the photometric data. Then, we built a second model with the transiting planet and a parametrisation of the 3-year signal as a cubic or a sinusoidal function. We find that a model that accounts for the 3-year signal is largely favored with a Bayes factor larger than 200. Comparing the cubic versus the sinusoid model, we find that the latter is slightly favoured with a Bayes factor of 5.

Finally, we looked at the potential modelling of the activity induced by the stellar rotation period at 16 days,
which could produce the third signal at 34 days seen in the periodogram analysis. To do so, we used our previous model
containing the transiting planet and a sinusoid function and we added a Gaussian process (GP) with a quasi-periodic kernel.
The maximum likelihood solution is shown in Appendix~\ref{fig:plot_maxL}. 
We find that this last model leads to a marginal improvement, compared to a model without a GP, with a Bayes factor of only 2.8.
We verified that the planetary parameters obtained are consistent with the model without a GP. In particular,
  the planetary mass is consistent well within $\rm 1\,\sigma$ between the models.
Hence we chose to move forward with a model without a GP which only accounts for the transiting planet along with the 3-year signal.

\subsection{Photometry and radial velocity analysis}
\label{sec:joint_fit}

We used the software package \texttt{juliet} \citep{espinoza_2019a} to model the photometric and radial velocity data of \ticA.
Using a Bayesian framework, \texttt{Juliet} allows one to model multi-planetary systems with a given number of
transiting and non-transiting objects. The planetary transits are modelled with \texttt{batman} \citep{kreidberg_2015}
and the Keplerian signals are modelled with \texttt{RadVel} \citep{fulton_2018a}. The stellar activity signals and instrument systematics
can be taken into account with parametric functions or Gaussian processes (e.g. \citealt{gibson_2014}).

We chose to only include TESS sector 31 in our analysis since this sector is the one displaying a transit event.
Similarly for the NGTS data, we included in the joint analysis only the nights which display a partial transit event.
We specified for the TESS and NGTS photometric filters two sets of quadratic limb-darkening parameters \citep{kipping_2013}.
The priors on the limb-darkening parameters were obtained with \texttt{LDCU}\footnote{\url{https://github.com/delinea/LDCU}} code.
\texttt{LDCU} is a modified version of the python routine implemented by \citet{espinoza_2015}, that computes the limb-darkening coefficients
and their corresponding uncertainties using a set of stellar intensity profiles accounting for the uncertainties on the stellar parameters.
The stellar intensity profiles are generated based on two libraries of synthetic stellar spectra: ATLAS \citep{kurucz_1979} and PHOENIX \citep{husser_2013a}.
We used the most conservative estimate from \texttt{LDCU} which is the Merged all results.

We also specified three sets of offset and jitter
terms for the TESS sector 31 and the two nights of NGTS data because the NGTS data were taken more than one year apart.
We did not include a dilution factor for either instrument as the TESS and NGTS apertures are not contaminated by other sources.
To correctly evaluate the baseline of the in-transit data in the TESS photometry, we chose to model the photometric variability with
a Gaussian process using a Mat\'{e}rn 3/2 kernel. We tested using a SHO kernel with broad log-uniform priors on the scale amplitude, angular frequency, and quality factor. There was no improvement in the fit as measured by the evidence value and the planetary parameters were all consistent within 1\,$\sigma$.

We modelled the planetary signal with the following parameters: orbital period, planet-to-star radius ratio,
mid-transit time, impact parameter, argument of periastron, eccentricity, and radial velocity semi-amplitude.
We chose to model the eccentricity and argument of periastron with the
$\rm \sqrt{e}cos(\mathsf{\omega})$ - $\rm \sqrt{e}sin(\mathsf{\omega})$ parametrization \citep{eastman_2013}.
On top of the reflex motion induced by the transiting planet, the radial velocity data show a long-term signal with a periodicity close to 3 years.
This signal is probably induced by the magnetic activity cycle (see Section~\ref{sec:3year-signal} for more details) and we modelled it with a circular orbit.
The data from the four spectrographs CHIRON, CORALIE, FEROS, and HARPS were modelled by specifying an independent set of offset and jitter per instrument.
Moreover, we chose to model the stellar density with a normal prior derived from the stellar analysis performed in Section~\ref{stellar-analysis}.
The priors on all fitted parameters are detailed in Table~\ref{table:prior-juliet}.
We used the nested sampling method \texttt{dynesty} \citep{speagle_2019} implemented in \texttt{juliet} with 2000 live points and stopped
sampling once the uncertainty on the log-evidence is smaller than 0.1.

\begin{figure}
  \includegraphics[width=\hsize]{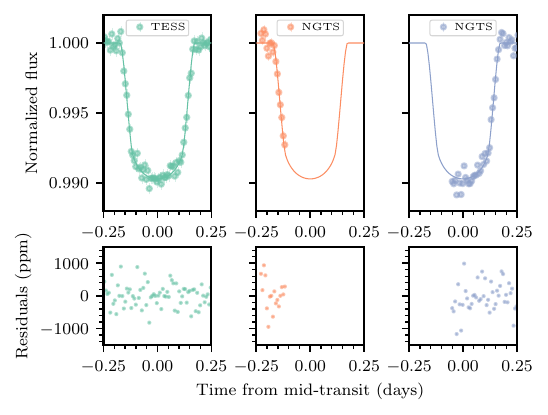}
  \caption{\textit{Top}: TESS and NGTS detrended photometric observations of \ticA\ (TESS in green, NGTS in orange and blue) with the common transit model (solid lines). NGTS data is binned at 10 minute cadence to match the TESS cadence. \textit{Bottom}: each panel shows the residuals between the transit model and the respective light curve.}
  \label{fig:lc_phase}
\end{figure}

\begin{table}[ht!]
  \caption{Fitted and derived parameters of \ticA\,b.}
  \label{table:system-parameters}
  \centering
  \begin{tabular}{l c}
    \hline
    \hline
    \noalign{\smallskip}
    Parameters                                 & \ticA\,b             \\
    \hline
    \noalign{\bigskip}
    \textit{Transiting planet}                          &                   \\
    \noalign{\smallskip}
    Orbital period (days)                      & $106.14468 ^{+0.00022} _{-0.00021}$ \\ 
    Time of transit $\rm T_{0}$ (days)         & $2459153.5934 ^{+0.0006} _{-0.0007}$ \\
    Radius ratio                               & $0.0967 ^{+0.0006} _{-0.0006}$ \\
    Impact parameter                           & $0.704 ^{+0.015} _{-0.016}$ \\
    $\rm \sqrt{e}cos(\mathsf{\omega})$         & $-0.02 ^{+0.06} _{-0.06}$ \\
    $\rm \sqrt{e}sin(\mathsf{\omega})$         & $-0.31 ^{+0.06} _{-0.04}$ \\
    Eccentricity                               & $0.098 ^{+0.028} _{-0.030}$  \\
    Argument of periastron                     & $266 ^{+12} _{-12}$           \\
    RV semi-amplitude ($\rm m\,s^{-1}$)            & $28.8 ^{+1.4} _{-1.5}$   \\

    \noalign{\bigskip}
    \textit{Host star and stellar activity cycle} &\\
    Stellar density ($\rm kg\,m^{-3}$)           & $1258 ^{+79} _{-80}$       \\ 
    Period activity cycle (days)              & $1140 ^{+82} _{-75}$ \\
    RV semi-amplitude ($\rm m\,s^{-1}$)   & $20.0 ^{+2.0} _{-1.9}$ \\

    \noalign{\bigskip}
    \textit{Derived parameters}                         &                  \\
    \noalign{\smallskip}
    Planetary radius ($R_{\rm J}$)             & $1.002 ^{+0.009} _{-0.009}$ \\
    Planetary radius ($R_{\oplus}$)            & $11.24 ^{+0.10} _{-0.10}$ \\
    Planetary mass ($M_{\rm J}$)               & $0.70 ^{+0.04} _{-0.04}$  \\
    Planetary mass ($M_{\oplus}$)              & $222 ^{+13} _{-13}$       \\
    Inclination (degrees)                      & $89.55 ^{+0.04} _{-0.04}$  \\
    Semi major axis (au)                       & $0.450 ^{+0.010} _{-0.010}$ \\
    Transit duration (hours)                   & $8.26 ^{+0.06} _{-0.06}$    \\
    Equilibrium temperature (K)                & $400 ^{+40} _{-70}$  \\

    \noalign{\bigskip}
    \textit{Instrumental parameters}                    &                  \\
    \noalign{\smallskip}
    TESS limb darkening q1                     & $0.322 ^{+0.034} _{-0.033}$   \\
    TESS limb darkening q2                     & $0.248 ^{+0.029} _{-0.026}$   \\
    NGTS limb darkening q1                     & $0.378 ^{+0.035} _{-0.032}$   \\
    NGTS limb darkening q2                     & $0.259 ^{+0.026} _{-0.029}$  \\
    TESS offset                                & $-0.00009 ^{+0.00009} _{-0.00010}$ \\
    TESS jitter (ppm)                          & $3.7 ^{+21.5} _{-3.3}$            \\
    NGTS 1 offset                              & $0.00113 ^{+0.00022} _{-0.00023}$   \\
    NGTS 1 jitter (ppm)                        & $5623 ^{+155} _{-154}$             \\
    NGTS 2 offset                              & $-0.00295 ^{+0.00008} _{-0.00009}$   \\
    NGTS 2 jitter (ppm)                        & $6522 ^{+112} _{-108}$                \\

    GP amplitude TESS  (relative flux)           & $0.00026 ^{+0.00008} _{-0.00005}$ \\
    GP time-scale TESS  (days)                   & $1.7 ^{+0.7} _{-0.5}$               \\

    CHIRON offset ($\rm m\,s^{-1}$)             & $-4.845 ^{+0.004} _{-0.004}$   \\
    CORALIE offset ($\rm m\,s^{-1}$)            & $-3.4151 ^{+0.0027} _{-0.0025}$ \\
    FEROS offset ($\rm m\,s^{-1}$)              & $-3.421 ^{+0.004} _{-0.005}$     \\
    HARPS offset ($\rm m\,s^{-1}$)              & $-3.3942 ^{+0.0020} _{-0.0020}$   \\ 
    CHIRON jitter ($\rm m\,s^{-1}$)             & $3 ^{+4} _{-2}$\\
    CORALIE jitter ($\rm m\,s^{-1}$)            & $6^{+4}_{-3}$     \\
    FEROS jitter ($\rm m\,s^{-1}$)              & $14^{+5}_{-4}$   \\
    HARPS jitter ($\rm m\,s^{-1}$)              & $7.8^{+1.2}_{-1.1}$   \\
    \hline
    
  \end{tabular}
\end{table}

\section{Results}
\label{results}

\subsection{Planetary and orbital parameters of TOI-2449\,b}

We jointly modelled the photometric and radial velocity data of \ticA\ and we find that this G0-star hosts a transiting planet
on a 106-day orbit. The planet \ticA\,b has the following characteristics : a mass of $\rm 0.70 ^{+0.05} _{-0.04}\,M_{J}$ and a radius of $\rm 1.001\pm0.009\,R_{J}$.
The planetary orbit has a semi-major axis of $0.449 ^{+0.011} _{-0.008}$ au and a small eccentricity of $0.098 ^{+0.028} _{-0.030}$.
The total transit duration is $\rm 8.26\pm0.06\,hours$.
The full transit of \ticA\,b observed with TESS at a 10-min cadence and the two partial transits observed with NGTS (one camera in 2021 and six cameras in 2022)
binned to a 10-min cadence are presented in Fig.~\ref{fig:lc_phase}. In Fig.~\ref{fig:tess_full_plot}, we show the full TESS light curve of Sector 31. After subtraction of the median model, the lightcurves are without structures and the residuals of the TESS data have a standard deviation of 408 ppm. We verified that the residuals follow a normal distribution by using a Kolmogorov-Smirnov test. We measured the standard deviation of the residuals as a function of the time bins (ranging from 3-hour to 10-minute binning) and we confirm that the residuals follow the same decreasing trend as samples taken from a normal distribution.
The residuals of the first NGTS observation, done with one camera and binned to 10 minutes, has a standard deviation of 460 ppm.
The residuals of the second NGTS observation, binned to 10 minutes, has a standard deviation which ranges from 980 to 1280 ppm for each camera,
once the data from the six cameras are combined the standard deviation is 450 ppm.

\begin{figure}
  \includegraphics[width=\hsize]{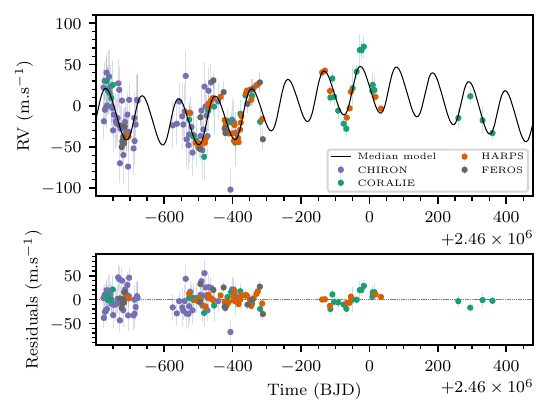}
  \caption{Radial velocity timeseries with median model (top panel) and residuals (bottom panel).}
  \label{fig:rv_time}
\end{figure}

\begin{figure}
  \includegraphics[width=\hsize]{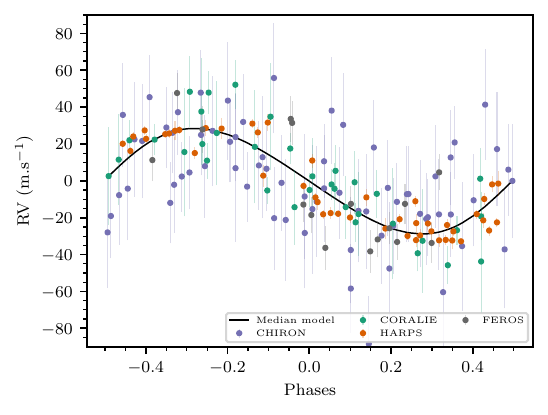}
  \caption{Phase folded radial velocity plot of \ticA\,b.}
  \label{fig:rv_phase}
\end{figure}

The radial velocity semi-amplitude induced by \ticA\,b is $\rm 28.8^{+1.8}_{-1.5}\,m\,s^{-1}$.
We present the radial velocity time series with the median model and the associated residuals in Fig.~\ref{fig:rv_time} and the phase-folded plot in Fig.~\ref{fig:rv_phase}.
The standard deviation of the residuals with respect to that median model are 24, 13, 8, and 17 $\rm m\,s^{-1}$ for CHIRON, CORALIE, HARPS, and FEROS respectively.
In comparison the average radial velocity errors are 27, 17, 4, and 10 for these instruments.
The HARPS residuals are larger that the typical radial velocity error of 4 $\rm m\,s^{-1}$.
This discrepancy is probably due to the additional 34-day signal which is not modelled in our final analysis.
For the CORALIE, CHIRON, and FEROS residuals, the HARPS data dominates the radial velocity part of the fit leading to a better fit for the data from other spectrographs.
Finally, we find that the long-term signal in the radial velocities has an periodicity of $\rm 1137 ^{+83} _{-71} \,days$.

We report the final parameters and their uncertainties for the planetary system around \ticA\ and the instrumental parameters in Table~\ref{table:system-parameters}. In Fig.~\ref{fig:corner_2449}, we present the posterior distributions for the parameters of the planet, the stellar density, and the stellar activity cycle.

\subsection{Interior modelling of TOI-2449\,b}
\label{sec:interior}

\ticA\,b is part of the growing sample of transiting warm Jupiters.
Assuming a Jupiter-like bond albedo (A=0.343) and full heat redistribution, \ticA\,b has an equilibrium temperature of $\rm 400^{+40}_{-70}\,K$. The upper and lower bounds were calculated with A=0 and A=0.686 respectively. At this equilibrium temperature, the atmosphere of this planet is not expected to be inflated.
Indeed \ticA\,b receives an insolation flux of $\rm 1.10^7\, erg\,s^{-1}\,cm^{-2}$, which is
lower than the usual threshold for inflation of $\rm 2.10^8\, erg\,s^{-1}\,cm^{-2}$ observed by \citet{miller_2011} and \citet{demory_2011}.
This insolation threshold corresponds to an equilibrium temperature of about 1000\,K.

Transiting warm Jupiters are ideal to study their internal composition. The link between the interior composition and the planetary size is more straightforward
because these planets are not subject to inflation mechanisms. The heavy element content is one parameter which can be well constrained by the interior modelling of gas giant planets.
For gas giant planets, the majority of the mass is made of hydrogen and helium and only a small fraction is made of other elements which we refer to as heavy elements or simply metals.
Heavy elements have a large impact on the planetary radius and increasing the heavy element content leads to smaller radii (e.g. \citealt{baraffe_2008,miller_2011}).

We used the planetary evolution code \texttt{completo} \citep{mordasini_2012} to build a grid of interior models for gas giant planets orbiting FGK and M stars. The planets are modelled as a core and an envelope coupled with a semi-gray atmospheric model. The planets have masses ranging from 15 $\rm M_{\oplus}$ to 13 $\rm M_{Jup}$ and semi-major axes ranging from 0.1 to 10 au. The core mass can vary from 0 to 10 $\rm M_{\oplus}$ and is made of iron and silicates, with an iron mass fraction of 33\%. The H/He envelope is modelled with the H/He equation of state (EoS) from \cite{chabrier_2021}. The heavier elements are assumed to be homogeneously mixed in the envelope and are modelled as water according to the AQUA EoS \citep{haldemann_2020}.
We did not consider inflation mechanisms, as it is unlikely to affect planets on such wide orbits (e.g. \citealt{demory_2011}).
The grid of evolution models was then coupled to a Bayesian inference model to retrieve the heavy element content of \ticA\,b.

In this framework, we used information from the photometric, radial velocity joint fit and the host star analysis to set normal priors on the planetary mass, stellar mass, and stellar age and to fix the semi-major axis to its median value. The prior on the metal fraction of the envelope ($\rm Z_{env}$) is a uniform prior between 0 and 0.9.
Given the planetary parameters, $\rm Z_{env}$, and stellar age, we calculated the corresponding evolution models. We then obtained the distribution of theoretical radii and interior compositions compatible with TOI-2449 b.

We find that \ticA\,b has an envelope marginally enriched in heavy elements by a fraction  $\rm Z_{env} = 0.03 ^{+0.03} _{-0.02}$.
The core mass is not well constrained by our retrieval ($\rm 3.1^{+3.2}_{-2.2} \,M_{\oplus}$).
By adding the mass of heavy elements in the envelope and the core mass, we find that the total heavy element mass is $\rm M_z = 11 ^{+6} _{-5} \,M_{\oplus}$.
Hence, the total heavy element mass fraction ($\rm M_z / M_p$) is  $\rm Z_p = 0.043 ^{+0.033} _{-0.024}$.
We estimated the overall stellar metallicity from the iron abundance as follows $\rm Z_{\star} = 0.0142 \times 10^{[Fe/H]}$ \citep{asplund_2009,miller_2011}
and obtained $\rm Z_{\star} = 0.0133 \pm 0.0012$.
The planet metal enrichment of \ticA\,b is $\rm Z_p / Z_{\star}  = 3.3^{+2.5}_{-1.8} $.
Due to its relatively low planetary mass compared to its radius, the Jupiter-sized planet \ticA\,b is consistent with no metal enrichment relative to its host star at 2$\rm \sigma$.
We note that several interior models exist to probe the composition of giant exoplanets (e.g. \citealt{muller_2020,acuna_2024,sur_2024}) and may lead to differences in the exact derived values of planet metallicity.

\section{Discussion}
\label{sec:discussion}

\begin{figure}[h]
  \includegraphics[width=0.95\hsize]{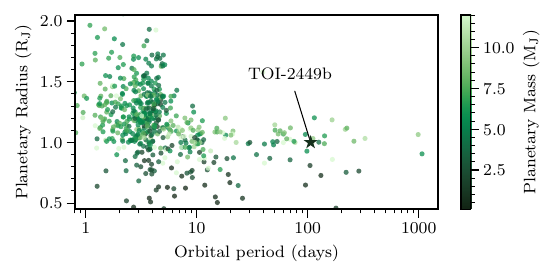}
  \caption{Planetary radius as a function of orbital period color-coded by the planetary mass for known transiting giant exoplanets.}
  \label{fig:radius_insolation}
\end{figure}

\begin{figure}[h]
  \includegraphics[width=0.95\hsize]{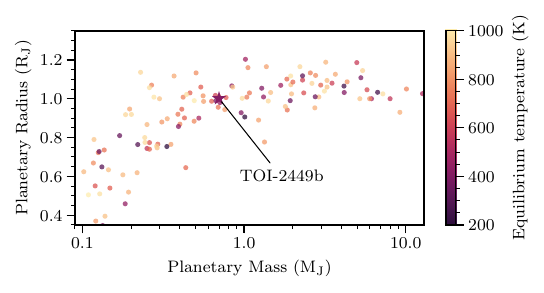}
  \caption{Radius-mass distribution color-coded by the equilibrium temperature for transiting warm giant exoplanets (Teq < 1000 K).}
  \label{fig:radius_mass}
\end{figure}

\ticA\,b joins the small population of well-characterised transiting planets orbiting a bright host star with a relatively long period of 106 days.
In Fig.~\ref{fig:radius_insolation}, we present \ticA\,b within the sample of transiting exoplanets taken from the TEPCat
catalog\footnote{\url{https://www.astro.keele.ac.uk/jkt/tepcat/}}. We selected exoplanets with planetary masses larger than $\rm 0.1\,M_{J}$
and uncertainty on their planetary mass and radius measurements smaller than 8\% and 20\% respectively.
While there is a growing sample of well-characterised transiting gas giant planets with orbital periods larger than 10 days,
only 23 planets have an orbital period larger than 100 days and 10 planets transit a bright host star (Vmag < 12).
Figure~\ref{fig:radius_mass} displays only the warm giant planets with an equilibrium temperature below 1000\,K.
This temperature threshold is commonly used to define the limit below which the planetary atmosphere is usually not inflated.
From Fig.~\ref{fig:radius_mass}, we notice that the warm giant planets do not have planetary radii above $\rm 1.2\,R_{J}$.

We compared TOI-2449 b with the outputs of the New Generation Planetary Population Synthesis (NGPPS) formation and evolution model \citep{emsenhuber_2021}
which is built on the core accretion paradigm. We used the NG76longshot population based on \citet{emsenhuber_2021a}
with a longer integration time of 100 Myr and the inclusion of realistic water phases (see \citealt{burn_2024} for details).
We observe that synthetic planets close to the mass, radius, and semi-major of TOI-2449 b form in the outer part of the disc between 3 and 6 au.
These planets first grow to about 10 $M_{\rm \oplus}$ at their formation location, and then migrate inwards while accreting large quantities
of gas and increasing their total mass.
TOI-2449 b contains about 11 $M_{\rm \oplus}$ of metals, a value which is in agreement with the closest synthetic planets in this population.
Hence, TOI-2449 b likely formed at a few aus from its host star, then migrated inward while essentially accreting gas to be found at its current location.
At this stage, we cannot pinpoint the migration pathway of TOI-2449 b. A measure of the projected spin-orbit angle of the system would help us distinguish between possible migration mechanisms such as disc migration or high-eccentricity migration.
The expected Rossiter-McLaughlin (RM) signal is about 35 $\rm m.s^{-1}$ (Eq 40, \citealt{winn_2010}).
Thanks to the v.sini of the host star, the RM effect is large and should be easily detectable with high-resolution spectrographs such as HARPS and ESPRESSO. Given the long transit duration of \ticA\,b, ground-based spectroscopic observations could be performed in two separate visits covering the first and second half of the transit.
In the next three years, there are two transit opportunities from Paranal, Chile.
The mid transit times in BJD are : $\rm 2461382.632 ^{+0.005} _{-0.004}$ (08.Dec.2026 03:10 UTC) and $\rm 2462125.644 ^{+0.006} _{-0.006}$ (20.Dec.2028 03:27 UTC)

Finally, atmospheric characterisation of such warm and temperate planets would be useful to understand the link between the well-studied hot Jupiters
and the gas giants of the Solar system.
We estimated the Transmission Spectrocopy Metric (TSM, \citealt{kempton_2018}) of \ticA\,b to be 38. The TSM value is low compared to the values for hot Jupiters. However, for a temperate giant planet, a TSM of around 30 is enough to be considered as one of the best targets to perform transmission spectroscopy (e.g. Kepler-16b, \citealt{hord_2024}). With an equilibrium temperature of about 400 K and a mass measured with a precision better than 10\%, \ticA\,b is a suitable target for transmission spectrocopy with JWST for the class of temperate giant planets.
In Section~\ref{sec:interior}, we estimated the planetary metal mass fraction Zp using a planetary evolution model.
From this metal mass fraction, we can infer metal abundance ratio Z:H (by number) following Equation 3 from \citet{thorngren_2019a}. 
We find a Z:H ratio of $\rm 6^{+5}_{-4}\times solar$. This value provides an upper limit on the atmospheric metallicity,
since Zp was derived assuming that the metals in the envelope are homogeneously mixed.

\begin{figure}[h]
  \includegraphics[width=\hsize]{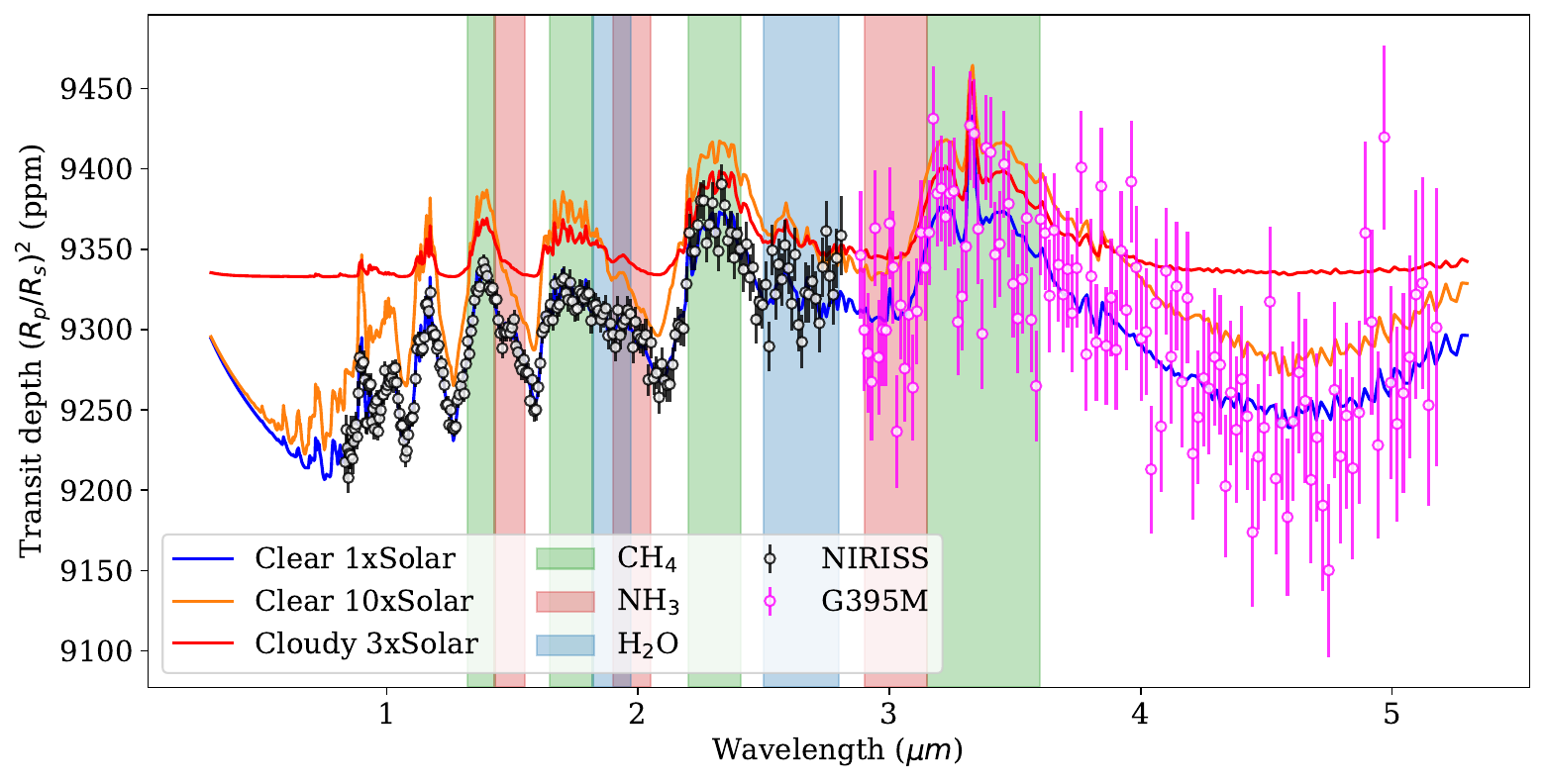}
  \caption{Synthetic transmission spectra of TOI-2449\,b from {\tt petitRADTRANS} \citep{molliere2019petitradtrans}, assuming equilibrium chemistry and an isothermal atmosphere. We use 1$\times$, 3$\times$, and 10$\times$ models with a solar C/O ratio, and include a grey cloud deck at 1 mbar for the cloudy model. Simulated flux measurements from NIRISS/SOSS and NIRSpec/G395M instrument are shown as black and pink points.}
  \label{fig:transmission_spectra}
\end{figure}

Atmospheric characterisation of temperate gas giants like TOI-2449\,b would allow us to build a connected understanding
of giant planet populations across the parameter space, combining with dynamical clues like eccentricity, obliquity,
and system architecture preserved in the orbits of these planets to infer the planets formation and migration history \citep{dawson_2018}.
Particularly notable at these temperatures, \ch{CH4} and \ch{NH3} are the equilibrium carriers of carbon and nitrogen \citep{fortney2020beyond},
compared to \ch{CO} and \ch{N2} for hot Jupiters.
Notably \ch{NH3} only appears in these cooler planets (<500 K), and unlike \ch{N2} can be detected with transmission spectroscopy.
This means that the C/N/O ratio can be uniquely measured, which can be more precisely related
to the planet's formation and migration history than C/O ratio alone \citep[e.g.,][]{piso2016role,cridland_2020}.
In Fig.~\ref{fig:transmission_spectra}, we present synthetic transmission spectra for TOI-2449\,b using
an equilibrium chemistry model and different metallicities within the range implied
by internal structure modelling using {\tt petitRADTRANS} \citep{molliere2019petitradtrans}.
Simulated observations from the NIRISS/SOSS and NIRSpec/G395M instruments \citep{batalha_2017} are displayed along the simulated spectra, showing detectable variations in particular for the \ch{CH4} features at 2.2 and \qty{3.3}{\micro\meter}, and \ch{NH3} feature at 1.5 and \qty{3}{\micro\meter}.

Furthermore, the atmospheres of temperate gas giants are predicted to host diverse disequilibrium chemistry processes, including condensation of both \ch{S8} and \ch{H2O} \citep{gao2017sulfur,mang2022microphysics}, photochemical destruction of \ch{NH3} and production of HCN \citep{hu2021photochemistry,ohno2023nitrogen2}, as well as chemical quenching due to vertical mixing \citep{fortney2020beyond,ohno2023nitrogen}. These processes are dependent on many factors, including the temperature, stellar host, metallicity, and gravity, making it important that we have a diverse sample of temperate gas giants for atmospheric characterisation. TOI-2449\,b joins TOI-2447\,b \citep{gill_2024}, NGTS-30\,b \citep{battley_2024} and TOI-2010\,b \citep{mann_2023} in a cluster of 400--500 K Jupiters around Sun-like hosts, covering a range of masses from 0.5--1.2 $\rm M_{J}$, and an excellent regime for understanding nitrogen chemistry.

\section{Conclusions}
\label{conclusion}

We report the discovery and characterisation of \ticA\,b / NGTS-36\,b, a 106-day warm giant exoplanet orbiting a G-type star. 
This planet was first identified as a single transit event in the TESS data.
Ground-based photometric follow-up with NGTS and radial velocity follow-up
solved the orbital period to 106.144~days corresponding to a semi-major axis of 0.449 au.
The planet has a mass of $\rm 0.70^{+0.05}_{-0.04}\,M_{J}$ and a radius of $\rm 1.001\pm0.009\,R_{J}$ and is on a slightly eccentric orbit.
We detect an additional 3-year periodic signal in the radial velocity data which is most likely the magnetic activity cycle of the star.
\ticA\,b / NGTS-36\,b joins the growing sample of transiting warm Jupiters orbiting relatively bright stars. 
Further observations of the RM effect produced by \ticA\,b / NGTS-36\,b
and atmospheric characterisation through transmission spectroscopy are achievable and will allow us to
learn more about the formation and evolution pathway of this system.

\subsection*{Acknowledgements}

This work has been carried out within the framework of the National Centre of Competence
in Research PlanetS supported by the Swiss National Science Foundation (SNSF) under grants 51NF40\_182901 and 51NF40\_205606.
The authors acknowledge the financial support of the SNSF.
This work is based on data collected under the NGTS project at the ESO Paranal Observatory. The NGTS facility is operated by a consortium institutes with support from the UK Science and Technology Facilities Council (STFC) under projects ST/M001962/1, ST/S002642/1 and ST/W003163/1. The contributions at the University of Warwick by SG, DB, PJW, RGW and DA have been supported by STFC through consolidated grants ST/P000495/1, ST/T000406/1 and ST/X001121/1.
KA acknowledges support from the SNSF under the Postdoc Mobility grant P500PT\_230225.
ML acknowledges support of the SNSF under grant number PCEFP2194576. 
This paper made use of data collected by the TESS mission and are publicly available from the Mikulski Archive for Space Telescopes (MAST) operated by the Space Telescope Science Institute (STScI). Funding for the TESS mission is provided by NASA’s Science Mission Directorate. Resources supporting this work were provided by the NASA High-End Computing (HEC) Program through the NASA Advanced Supercomputing (NAS) Division at Ames Research Center for the production of the SPOC data products.
KAC acknowledges support from the TESS mission via subaward s3449 from MIT.
R.B. acknowledges support from FONDECYT Project 1241963 and from ANID -- Millennium  Science  Initiative -- ICN12\_009. This work was funded by the Data Observatory Foundation. 
A.De. acknowledges financial support from the Swiss National Science Foundation (SNSF) for project 200021\_200726.
A.J. acknowledges support from ANID -- Millennium Science Initiative -- ICN12\_009, AIM23-0001 and from FONDECYT project 1210718.
M.T.P. acknowledges support from Fondecyt-ANID fellowship no. 3210253 and ASTRON-0037.
A.P acknowledges support from the Unidad de Excelencia María de Maeztu CEX2020-001058-M programme and from the Generalitat de Catalunya/CERCA.
The results reported herein benefitted from collaborations and/or information exchange within NASA’s Nexus for Exoplanet System Science (NExSS) research coordination network sponsored by NASA’s Science Mission Directorate under Agreement No. 80NSSC21K0593 for the program ``Alien Earths".
GV has received support from the European Research Council (ERC) under the European Union’s Horizon 2020 research and innovation programme (Grant Agreement No. 947660).
Some of the observations in this paper made use of the High-Resolution Imaging instrument Zorro and were obtained under Gemini LLP Proposal Number: GN/S-2021A-LP-105. Zorro was funded by the NASA Exoplanet Exploration Program and built at the NASA Ames Research Center by Steve B. Howell, Nic Scott, Elliott P. Horch, and Emmett Quigley. Zorro was mounted on the Gemini South telescope of the international Gemini Observatory, a program of NSF’s OIR Lab, which is managed by the Association of Universities for Research in Astronomy (AURA) under a cooperative agreement with the National Science Foundation. on behalf of the Gemini partnership: the National Science Foundation (United States), National Research Council (Canada), Agencia Nacional de Investigaci\'{o}n y Desarrollo (Chile), Ministerio de Ciencia, Tecnolog\'{i}a e Innovaci\'{o}n (Argentina), Ministério da Ci\^{e}ncia, Tecnologia, Inovaç\~{o}es e Comunicaç\~{o}es (Brazil), and Korea Astronomy and Space Science Institute (Republic of Korea).
C.A.W. would like to acknowledge support from the UK Science and Technology Facilities Council (STFC, grant number ST/X00094X/1).
JSJ greatfully acknowledges support by FONDECYT grant 1240738 and from the ANID BASAL project FB210003.
The contributions at the Mullard Space Science Laboratory by EMB have been supported by STFC through the consolidated grant ST/W001136/1.
TR is supported by an STFC studentship.

\bibliographystyle{aa}
\bibliography{PhotometryTransit.bib,extra.bib}

\begin{appendix}
\onecolumn
  \section{TESS target pixel file}
  \label{appendix:tpf_plot}
  
  \begin{figure*}[hbt!]  %[h!]
    \centering
    \includegraphics[width=0.5\hsize]{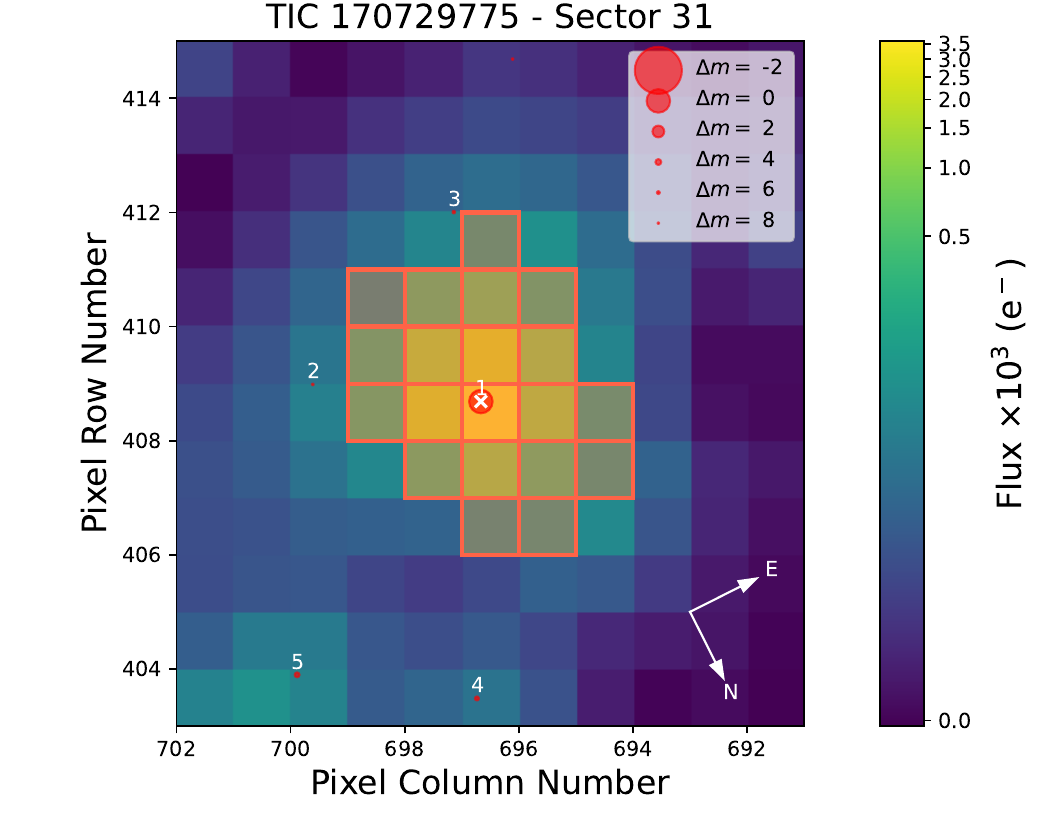}
    \caption{TESS target pixel file of \ticA\ with its aperture for the sector 31 plotted with the \textit{Gaia} EDR3 sources.}
    \label{fig:tess_tpf}
  \end{figure*}

  \section{NGTS observations}
  \label{appendix:ngts_plot}

  \begin{figure*}[hbt!]
  \includegraphics[]{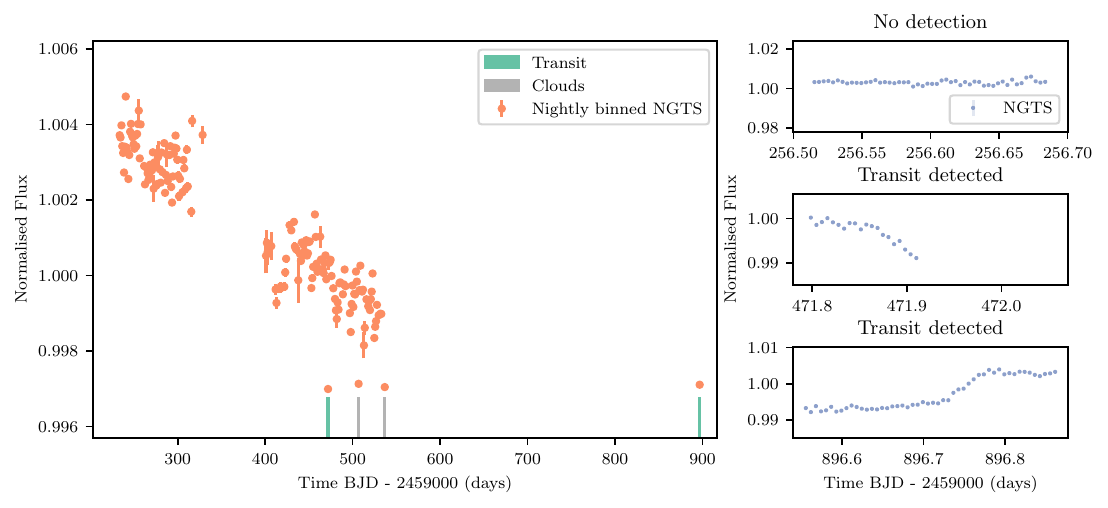}
  \caption{NGTS flux time series of TOI-2449. Left panel: Nightly binned NGTS data from January 2021 to November 2022 (orange dots). Right panel: Example of NGTS observations (blue dots) during a clear night with no detection (binned to 5 minutes; top), a clear night with a transit detection (binned to 10 minutes; middle and bottom).}
  \label{fig:all_ngts}
  \end{figure*}
  \clearpage

  \section{Imaging data}
  \label{appendix:imaging_plot}
  
  \begin{figure}[hbt!]
  \centering
  \includegraphics[width=0.45\hsize]{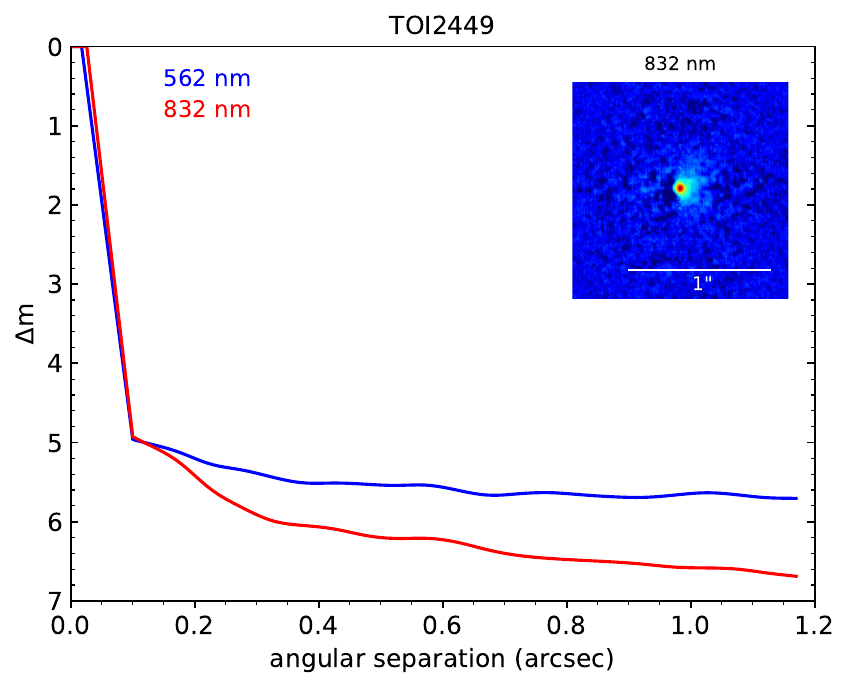}
  \caption{Speckle imaging contrast curves ($5\,\sigma$) in both filters as a function of the angular separation out to 1.2 arcsec from \ticA. The inset shows the speckle reconstructed 832 nm image with a 1 arcsec scale bar.}
  \label{fig:zorro}
  \end{figure}

  \begin{figure}[hbt!]
  \centering
  \includegraphics[width=0.45\hsize]{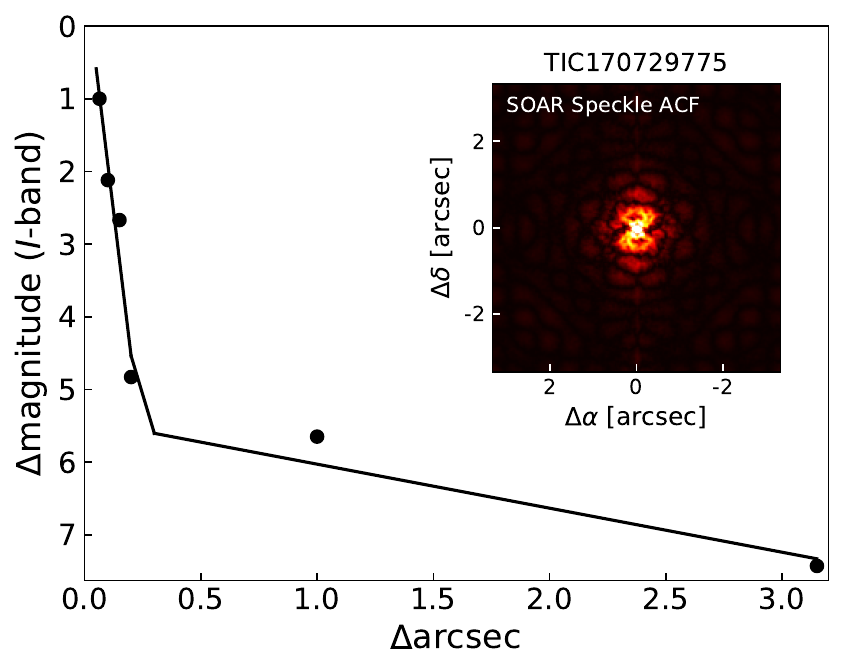}
  \caption{Detection sensitivity ($5\,\sigma$) and speckle auto-correlation functions from the SOAR observation of \ticA.}
  \label{fig:soar}
  \end{figure}
  \clearpage

  \section{Radial velocity modelling priors}
  \label{appendix:prior_rv}

  \begin{table*}[hbt!]
  \caption{Priors for the first radial velocity fit with a maximum of three planets.}
  \label{table:prior-kima}
  \centering
  \begin{tabular}{l l l}
    \hline
    \hline
    \noalign{\smallskip}
    Parameters                                 & Distribution     & Value \\
    \hline
    \noalign{\smallskip}
    Number of planets                          & Uniform          & (0, 3)   \\
    Orbital period (days)                      & LogUniform       & (1, 3000)       \\
    Semi-amplitude ($\rm ms^{-1}$)             & LogUniform       & (1, 200)       \\ 
    Eccentricity                               & Kumaraswamy      & (0.867, 3.03)       \\
    Mean anomaly                               & Uniform          & (-$\pi$, $\pi$)       \\
    Argument of periastron                     & Uniform          & (0, 2$\pi$)       \\
    Jitter ($\rm ms^{-1}$)                     & LogUniform       & (1, 200)       \\
    \hline
  \end{tabular}
  \end{table*}

  \section{Radial velocity fit with a GP}
  \label{appendix:gp}
  \begin{figure*}[hbt!]
    \centering
      \includegraphics[width=0.65\hsize]{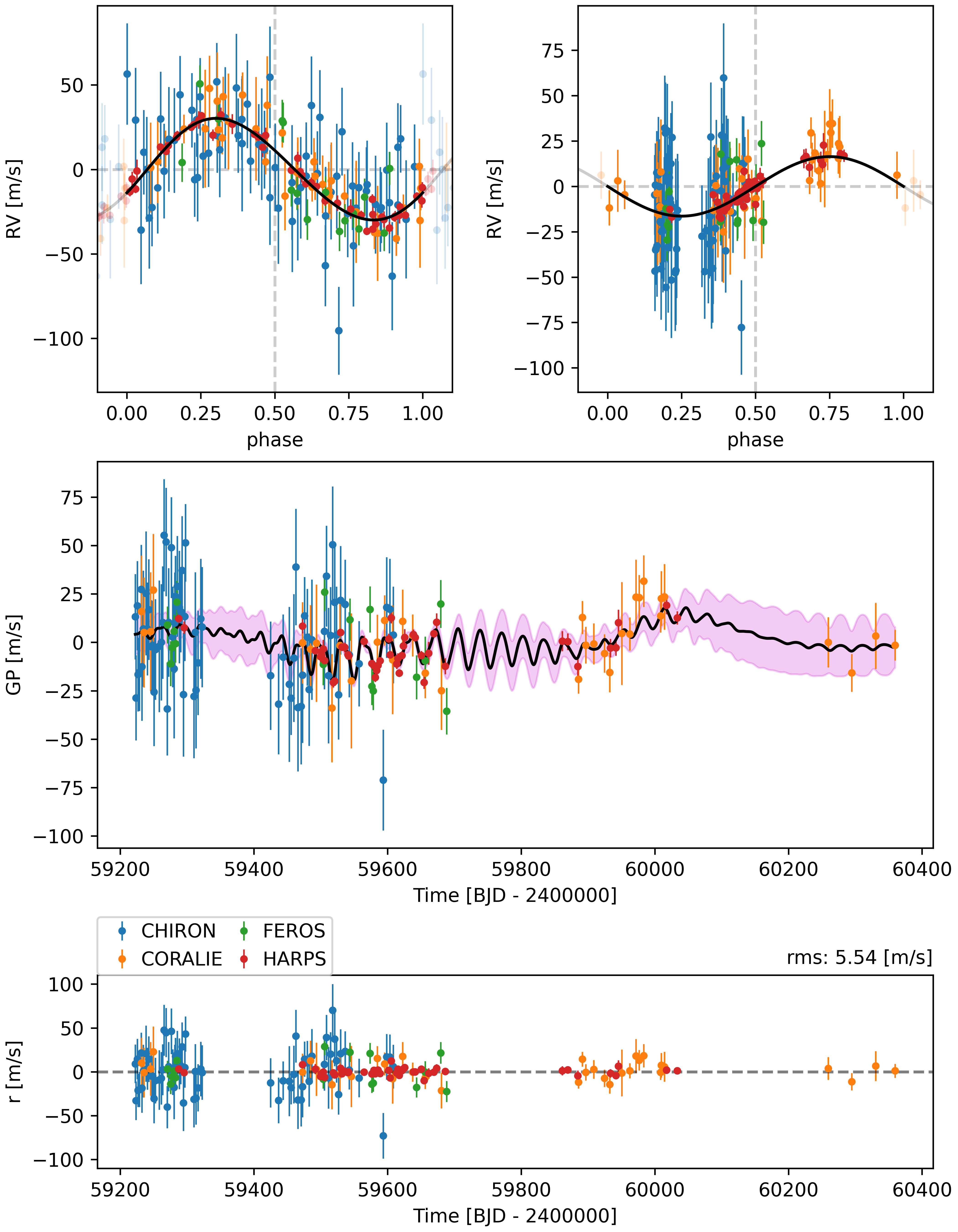}
    \caption{\textit{Top :} Radial velocity phase-folded plot of \ticA\,b (right) and the 3-year activity cycle modelled as a sinusoid (left) for the maximum likelihood solution.
      \textit{Middle :} Radial velocity dataset with the corresponding GP model (black).
    \textit{Bottom :} Radial velocity residuals.}
    \label{fig:plot_maxL}
    \end{figure*}
    \clearpage

  \section{Joint modelling priors}
  \label{appendix:prior_juliet}

  \begin{table*}[hbt!]
  \caption{Prior for the joint modelling of photometric and radial velocity data.
    The normal distribution is defined by its mean and variance. The uniform and log uniform distributions are defined by their lower and upper bounds. The priors on the radial velocity offsets and jitters are identical for CHIRON, CORALIE, FEROS, and HARPS.}
  \label{table:prior-juliet}
  \centering
  \begin{tabular}{l l c}
    \hline
    \hline
    \noalign{\smallskip}
    Parameters                                 & Distribution     & \ticA\\
    \hline
    \noalign{\bigskip}
    \textit{Transiting planet} &\\
    \noalign{\smallskip}
    Orbital period (days)                      & Uniform          & (30, 150)\\
    Time of transit $\rm T_{0}$ (days)         & Uniform          & (2459153.4, 2459153.8)\\
    Radius ratio                               & Uniform          & (0, 1)\\
    Impact parameter                           & Uniform          & (0, 1) \\
    $\rm \sqrt{e}cos(\mathsf{\omega})$         & Uniform          & (-1, 1) \\
    $\rm \sqrt{e}sin(\mathsf{\omega})$         & Uniform          & (-1, 1) \\
    RV semi-amplitude ($\rm kms^{-1}$)          & Uniform         & (0, 100)\\

    \noalign{\bigskip}
    \textit{Host star} &\\
    \noalign{\smallskip}
    Stellar density ($\rm kg.m^{-3}$)           & Normal           & (1260, 80)\\ 

    \noalign{\bigskip}
    \textit{Instruments}                    &                  \\
    \noalign{\smallskip}
    TESS limb darkening q1                     & Normal           & (0.32, 0.04)\\
    TESS limb darkening q2                     & Normal           & (0.24, 0.03)\\
    NGTS limb darkening q1                     & Uniform          & (0.37, 0.04) \\
    NGTS limb darkening q2                     & Uniform          & (0.26, 0.03) \\
    TESS offsets                               & Normal           & (0, 0.01)\\
    TESS jitters (ppm)                         & LogUniform       & (0.1, 1000) \\
    NGTS offsets                               & Normal           & (0, 0.01) \\
    NGTS jitters (ppm)                         & LogUniform       & (0.1, 1000) \\

    GP amplitude TESS (relative flux)           & LogUniform     & (1e-06, 100.0)\\
    GP time-scale TESS (days)                   & LogUniform     & (0.001, 100.0)\\

    Spectrograph offsets ($\rm kms^{-1}$)       & Uniform         & (-100, 100) \\ 
    Spectrograph jitters ($\rm kms^{-1}$)       & LogUniform      & (0.001, 0.2)\\

    \noalign{\bigskip}
    \textit{Stellar activity cycle} &\\
    \noalign{\smallskip}
    Orbital period (days)                      & LogUniform       & (1, 3000) \\
    Time of transit $\rm T_{0}$ (days)         & Uniform          & (2458745.0, 2459745.0)\\
    Eccentricity                               & Fixed            & 0           \\
    Argument of periastron                     & Fixed            & 90           \\
    RV semi-amplitude ($\rm kms^{-1}$)          & Uniform         & (0, 100)\\

    \hline
  \end{tabular}
  \end{table*}
  \clearpage

  \section{TESS observations}
\label{appendix:tess_plot}

\begin{figure*}[hbt!]
  \centering
    \includegraphics[width=0.65\hsize]{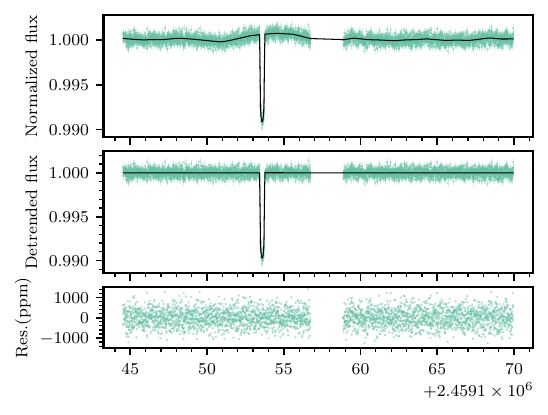}
    \caption{TESS photometric observations (green dots) with the median transit model (black line). Top : PDCSAP TESS light curves. Middle : Detrended TESS light curves. Bottom : Residuals between the median model and the TESS light curve. }
    \label{fig:tess_full_plot}
\end{figure*}
\clearpage

  \section{Joint modelling posterior distributions}
  \label{appendix:corner_juliet}

  \begin{figure*}[hbt!]
    \includegraphics[width=0.95\hsize]{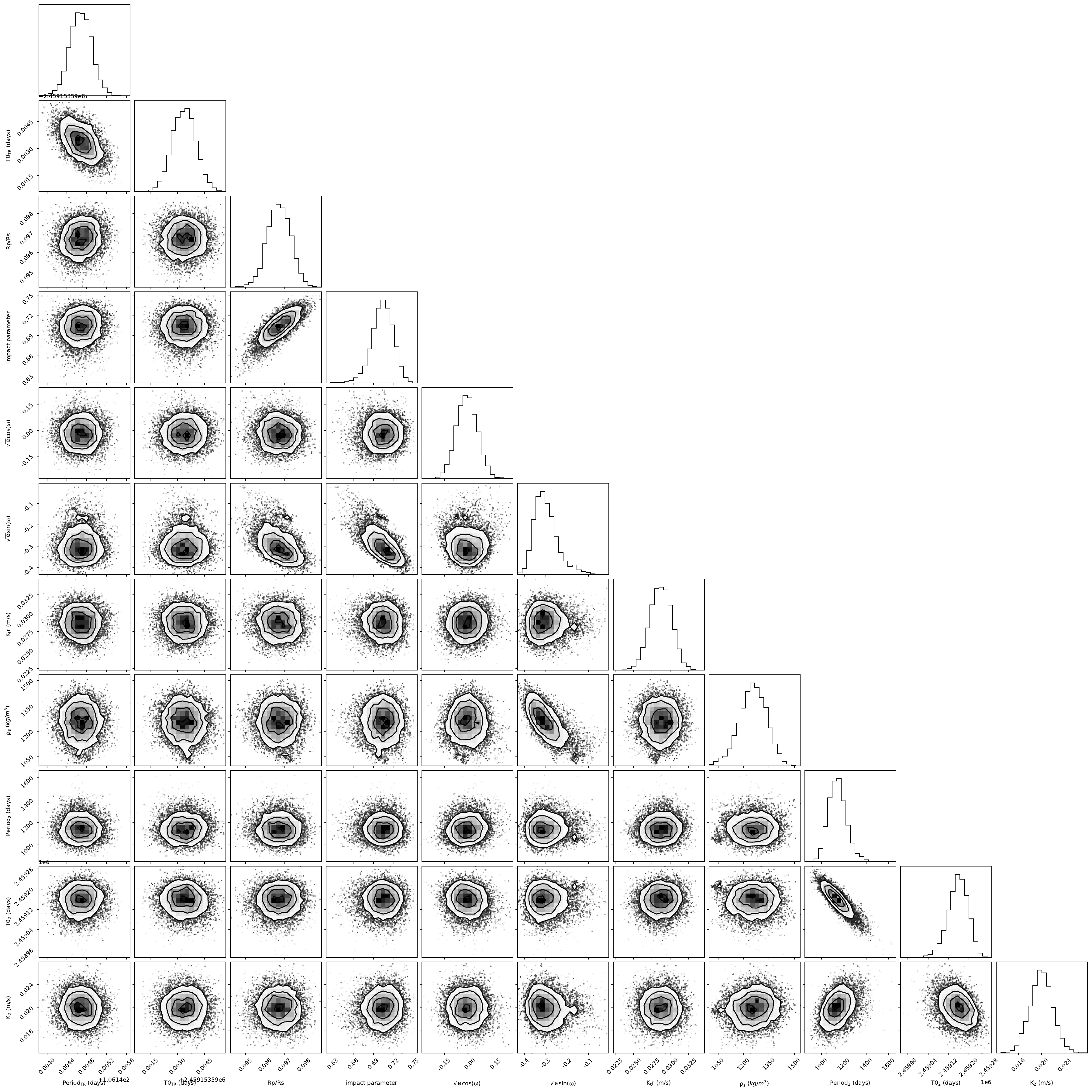}
    \caption{Posterior distributions of fitted parameters for \ticA\,b along with the stellar density ($\rm \rho_S$) and radial velocity semi-amplitude ($\rm K_{TR}$).
      The 3-year RV variation is parametrized with $\rm period_2, T0_2,$ and $\rm K_2$.}
    \label{fig:corner_2449}
  \end{figure*}
  \clearpage

\end{appendix}

\vfill
\eject
\end{document}